\documentclass[unnumsec,webpdf,contemporary,large]{template}%

\onecolumn
\usepackage{fancyhdr,amsmath,amssymb, amsthm,mathtools}
\usepackage{bbm,bm}
\usepackage{graphicx}
\usepackage{algorithm}
\usepackage[export]{adjustbox}

\graphicspath{{Fig/}}

\theoremstyle{thmstyleone}

\theoremstyle{thmstyletwo}

\theoremstyle{thmstylethree}

\newtheorem{Remark}{\bf Remark}

\newcommand{\enquote}{}
\newcommand{\bY}{ {\bf Y} }
\newcommand{\bPi}{ {\boldsymbol \Pi} }
\newcommand{\bz}{ {\bf z} }
\newcommand{\bLambda}{ {\boldsymbol \Lambda} }
\newcommand{\bc}{ {\bf c} }
\newcommand{\bX}{ {\bf X} }
\newcommand{\bW}{ {\bf W} }
\newcommand{\bZ}{ {\bf Z} }
\newcommand{\bS}{ {\bf S} }
\newcommand{\btheta}{ {\boldsymbol \theta} }
\newcommand{\balpha}{ {\boldsymbol \alpha} }
\newcommand\scalemath[2]{\scalebox{#1}{\mbox{\ensuremath{\displaystyle #2}}}}

\begin{document}

\journaltitle{Zero--Inflated Stochastic Block Modeling of Efficiency--Security  Tradeoffs in Weighted Criminal Networks}
\copyrightyear{2022}
\pubyear{2019}
\appnotes{$_.$}

\firstpage{1}

\title[Zero--Inflated Stochastic Block Modeling of Efficiency--Security  Tradeoffs in Weighted Criminal Networks]{\fontsize{16}{16}\selectfont Zero--Inflated Stochastic Block Modeling of Efficiency--Security  Tradeoffs in Weighted Criminal Networks}

\author[1]{Chaoyi Lu}
\author[2]{Daniele Durante$\ast$}
\author[1]{Nial Friel}

\authormark{Lu, Durante and Friel}

\address[1]{\orgdiv{School of Mathematics and Statistics}, \orgname{University College Dublin}, \country{Ireland}}
\address[2]{\orgdiv{Department of Decision Sciences and Bocconi Institute for Data Science and Analytics}, \orgname{Bocconi University},  \country{Italy}}

\corresp[$\ast$]{Corresponding author. \href{email: daniele.durante@unibocconi.it}{daniele.durante@unibocconi.it}}

\abstract{Criminal networks arise from the unique attempt to balance a need of establishing frequent ties among affiliates to~facilitate the coordination of illegal activities, with the necessity to sparsify the overall connectivity architecture to hide from law enforcement. This efficiency--security tradeoff is also combined with the creation of groups of redundant criminals that exhibit similar connectivity patterns, thus guaranteeing resilient network architectures. State--of--the--art models for such data are not designed to infer these unique structures. In contrast to such solutions we develop a computationally--tractable Bayesian zero--inflated Poisson stochastic block model (ZIP--SBM), which identifies groups of redundant criminals with similar connectivity patterns, and infers both overt and covert block interactions within and across such groups. This is accomplished by modeling weighted ties (corresponding to counts of interactions among pairs of criminals) via zero-inflated Poisson distributions with block--specific parameters that quantify complex patterns in the excess of zero ties~in each block (security) relative to the distribution of the observed weighted ties within that block (efficiency). The performance of the ZIP--SBM is illustrated in simulations and in a study of summits co--attendances in a complex Mafia organization, where we unveil efficiency--security structures adopted by the criminal organization that were hidden to previous analyses.}

\maketitle

\vspace{10pt}
\fontsize{10}{12}\selectfont

\section{\Large 1. Introduction}\label{sec_1}
The {\em ``EU Serious and Organised Crime Threat Assessment"} report recently  released by Europol in 2021~defines modern criminal networks as complex systems of interactions with incomplete information and a multifaceted combination of hierarchical overt and covert architectures. The challenges posed by these data incompletenesses and complexities undermine the effectiveness of law--enforcement policies, and present obstacles toward expanding knowledge on efficiency--security tradeoffs \citep{morselli2007efficiency}  of criminal organizations from the analysis of the interactions among the suspects observed during investigations \citep{lindquist2019crime,faust2019social,campana2022studying,diviak2022key}. Although substantial advancements have been made over the past thirty years in addressing these goals \citep[e.g.,][]{sparrow1991application,klerks2004network,von2006interdisciplinary,morselli2009inside,papachristos2014network,calderoni2017communities,bright2022}, as organized crime evolves towards more nuanced structures, the ability of current solutions to infer more complex architectures is undermined by a general reliance in criminology on overly--simplified representations and an overarching focus on descriptive analyses.

Recalling state--of--the--art reviews in the field \citep[e.g.,][]{campana2016explaining,lindquist2019crime,faust2019social,campana2022studying,diviak2022key}, there are at least two fundamental challenges which hinder advancements in the analysis of modern criminal networks. First, the covert~nature of criminal organizations implies an excess of zero ties in the observed criminal network, which yield a partial view of the actual connectivity architecture underlying the illicit organization. Second,~these zero ties are not randomly located, but often arise as the result of specific, yet unknown, secrecy and redundancy strategies to address efficiency--security tradeoffs \citep[][]{morselli2007efficiency,catino2015mafia,Bouchard,cavallaro2020disrupting}. As a consequence, these strategies lead to unique group structures formed by redundant criminals, along with complex block--interactions among such groups which incorporate a combination of dense community patterns, core--periphery structures~and weakly assortative modules  in both the overt and~covert connectivity architectures. This means that, if properly modeled, every block has the potential to disentangle both efficiency and security structures by studying the excess of zero ties within such a block (security) relative to the distribution of the observed weighted interactions among  criminals belonging to the two groups which identify the block (efficiency). 

This article is motivated by the above intuition and aims at translating the aforementioned challenges into opportunities for inferring more nuanced and yet--unaddressed efficiency and security architectures of criminal organizations rooted within the complex  interactions among the corresponding members. This is accomplished through the development of a new Bayesian zero--inflated Poisson stochastic block model (ZIP--SBM) for weighted, yet sparse, criminal networks which combines (i) species sampling processes \citep[e.g.,][]{de_2013,Gnedin2010} to rigorously characterize the mechanisms of redundant groups formation and criminals affiliation to these underlying groups,  (ii)  stochastic block models \citep[e.g.,][]{holland1983stochastic,nowicki2001estimation}  to define~the observed network as a function  of the criminals' allocations to the different groups and flexible block interactions among such groups  and, finally, (iii) zero--inflated Poisson distributions \citep[][]{lambert1992zero,ghosh2006bayesian}~to infer {\em unusual} excess of zeros in the distribution of the weighted ties within the blocks defining the different~pairs of groups. Such a model crucially exploits structural and regular equivalence patterns inherent to known endogenous and exogenous redundancies in organized crime  \citep[][]{sparrow1991application} to express the observed network  as a combination of two underlying ones. The first reconstructs, for each pair of groups, the actual strength of the ties among~the allocated criminals (efficiency), whereas the second could unveil the propensity of the illicit organization to either obscure or not these ties within the block corresponding to that pair of groups (security). Section~\ref{sec_1.0} clarifies~the methodological and applied advancements of the proposed approach.

\vspace{5pt}
\subsection{\large 1.1. Relevant literature}\label{sec_1.0}
\vspace{2pt}
Although there has been a recent adoption of sophisticated statistical methods to study criminal~networks \citep[e.g.,][]{malm2017more,charette2017network,calderoni2017communities,bright2019illicit,diviak2019structure,gollini2020modelling,cavallaro2020disrupting,legramanti2022extended},  none of the currently--available solutions provides a generative model that can flexibly incorporate, and~infer, core structures of illicit organizations not only in the strength of the observed ties among criminals, but also in systematic sparsity patterns, to ultimately unveil the nature of the efficiency--security~tradeoffs. In fact, popular link--prediction methods mainly rely on  descriptive solutions applied to a~dichotomized version of the observed weighted criminal networks \citep[][]{berlusconi2016link,calderoni2020robust}. These strategies lack a model--based perspective that would enable inference, uncertainty quantification and inclusion of those  measurement errors which may occur in investigations. Moreover, the loss~of information arising from dichotomization crucially fails to infer efficiency structures encoded in~the~strength~of~the weighted ties --- often corresponding to the~counts of interactions among pairs of criminals --- and hinders~the possibility to unveil more nuanced security architectures from the analysis of {\em unusual} patterns of zero connections relative to the distribution of  weighted ties.

A noteworthy attempt to move towards more structured model--based representations of the complex group interactions in modern criminal networks can be found in the extended stochastic block model of \citet{legramanti2022extended}. Albeit providing important advancements in inference on redundancy patterns relative to previous studies relying on community detection algorithms \citep[][]{girvan2002community,newman2006modularity,blondel2008fast} and spectral clustering methods \citep{von2007tutorial}, this perspective still focuses on dichotomized versions of weighted criminal networks. Hence, it faces the same conceptual barriers of link--prediction methods when the goal is to  disentangle and quantify security architectures from efficiency structures. In fact, in order to~infer an excess of zero ties pointing towards systematic obscuration mechanisms it is necessary to possess a benchmark distribution for the weighted interactions which allows one to quantify the extent to which the total~number~of~observed~zero connections is {\em unusual} relative to those expected under such a distribution. As clarified in Section~\ref{sec_2}, the proposed ZIP--SBM addresses these challenges via a novel generalization of  extended stochastic block models \citep{legramanti2022extended} in the context of weighted and sparse criminal networks where the focus is to quantify efficiency--security tradeoff architectures. This is accomplished by avoiding data dichotomization prior to statistical modeling while relying on block--specific zero--inflated Poisson distributions --- rather than Bernoulli ones ---  for the ties among groups of redundant criminals. 

Although the potentials of related constructions have been never explored in the context of criminal network analysis, from a methodological perspective there has been some research involving stochastic block models in combination with zero--inflations \citep{aicher2015learning,mariadassou2015convergence,ng2021weighted}. The proposed ZIP--SBM yields key advancements relative to these and other contributions. In fact, these formulations address the simpler, and very different,~settings which either consider all zero ties to denote a truly observed non--existing interaction or assume knowledge about which zero ties are associated to a truly non--existing interaction and which simply denote the lack of knowledge about the value of such an interaction. This information is~not available in commonly--analyzed criminal networks. In fact, the proposed ZIP--SBM aims to uncover this type of information rather than presuming it. In addition, the above procedures rely on multinomial~distributions for the group allocations which imply that the total number of modules underlying the criminal network is finite and pre--specified. In fact, such a quantity is unknown in criminal network studies and, hence, it is conceptually and practically useful to incorporate uncertainty also on the total number of groups and learn it as part of the inference process. As detailed in Section~\ref{sec_2.2},~the ZIP--SBM  addresses this aspect via a Gnedin process prior \citep{Gnedin2010} for the allocation of redundant criminals to groups \citep{legramanti2022extended}. Besides incorporating uncertainty in the~total number of modules within the network, this prior also yields a natural characterization for the  affiliation mechanism of criminals to redundant groups via a scheme that depends both on the size of the criminal network and also on the current number, dimension and composition of such groups. 

From a more general perspective, the proposed ZIP--SBM has also connections with a relevant line of~research that  considers the observed network as a corrupted measurement~of~a~``true''~underlying~one,~and~aims~at~recovering such an underlying network, while providing inference on its structures; see e.g., \citet{chatte2015}, \citet{priebe2015}, \citet{young2020} and the references therein. Recasting~our~proposal within this framework, the  ``true''  underlying network could be regarded as the actual count of interactions among pairs of criminals (efficiency), for which we observe a corrupted  version due to the zero inflation (possibly operated by a security strategy). However, unlike \citet{chatte2015} and \citet{young2020}, ZIP--SBM crucially includes and learns block structures among criminals both in the ``true''  underlying network and also in the mechanism yielding the corrupted measurements, while allowing the model parameters to~change across such blocks. These block structures are inherent to criminal networks and are also of key interest in criminology, thus motiving a focus on those models that can incorporate these structures. To this end, the contribution by \citet{priebe2015} is more in line with the scope of  ZIP--SBM, in that it also accounts for group structures among nodes in networks observed with errors. However,  \citet{priebe2015} focus on the simpler node classification problem in which the goal is to identify the group membership~of~a single distinguished node, assuming that those of all the others are already known.~This is not the case in criminal networks, where the group allocation defining redundancy patterns is unknown for all the criminals~and must be inferred from a network with contamination of zeros. 

Albeit providing a sophisticated representation of criminal networks, the proposed ZIP--SBM  is amenable to tractable posterior inference for the allocation of criminals to groups and for the parameters of the zero--inflated Poisson distribution within each block. This is accomplished~via a data--augmentation collapsed Gibbs--sampler derived in Section~\ref{sec_3}. The output of~this algorithm yields highly accurate reconstructions of redundant patterns and efficiency--security architectures both in extensive simulation analyses and also in a study of the complex Mafia network reconstructed from the judicial records of ``Operazione {\em Infinito}'' \citep[e.g.,][]{calderoni2017communities,legramanti2022extended}; see Section~\ref{sec_1.1}  for details about the motivating application. These analyses~are illustrated within Sections~\ref{sec_4}--\ref{sec_5}, respectively, and clarify the potential of a ZIP--SBM in uncovering relevant structures that state--of--the--art methods cannot unveil. For instance, unlike the analyses of the {\em Infinito} network in \citet{calderoni2017communities} and \citet{legramanti2022extended}, the proposed ZIP--SBM unveils a yet--undiscovered combination of communities and core--periphery architectures, in both~overt and covert connectivity patterns. While moving~from the periphery to the core, such structures~are characterized by a reduced redundancy combined with a tendency of establishing increasingly strong ties and a preference to progressively sparsify these connections to guarantee security of those criminals at the top of the pyramid. We refer the reader to Section~\ref{sec_6} for a final discussion and future  directions.

\vspace{5pt}

\subsection{\large 1.2. The 'Ndrangheta network from ``Operazione Infinito''}\label{sec_1.1}
\vspace{2pt}

\begin{figure}[t!]
\centering
    \includegraphics[trim=11cm 0cm 0.3cm 0cm,clip,width=0.85\textwidth]{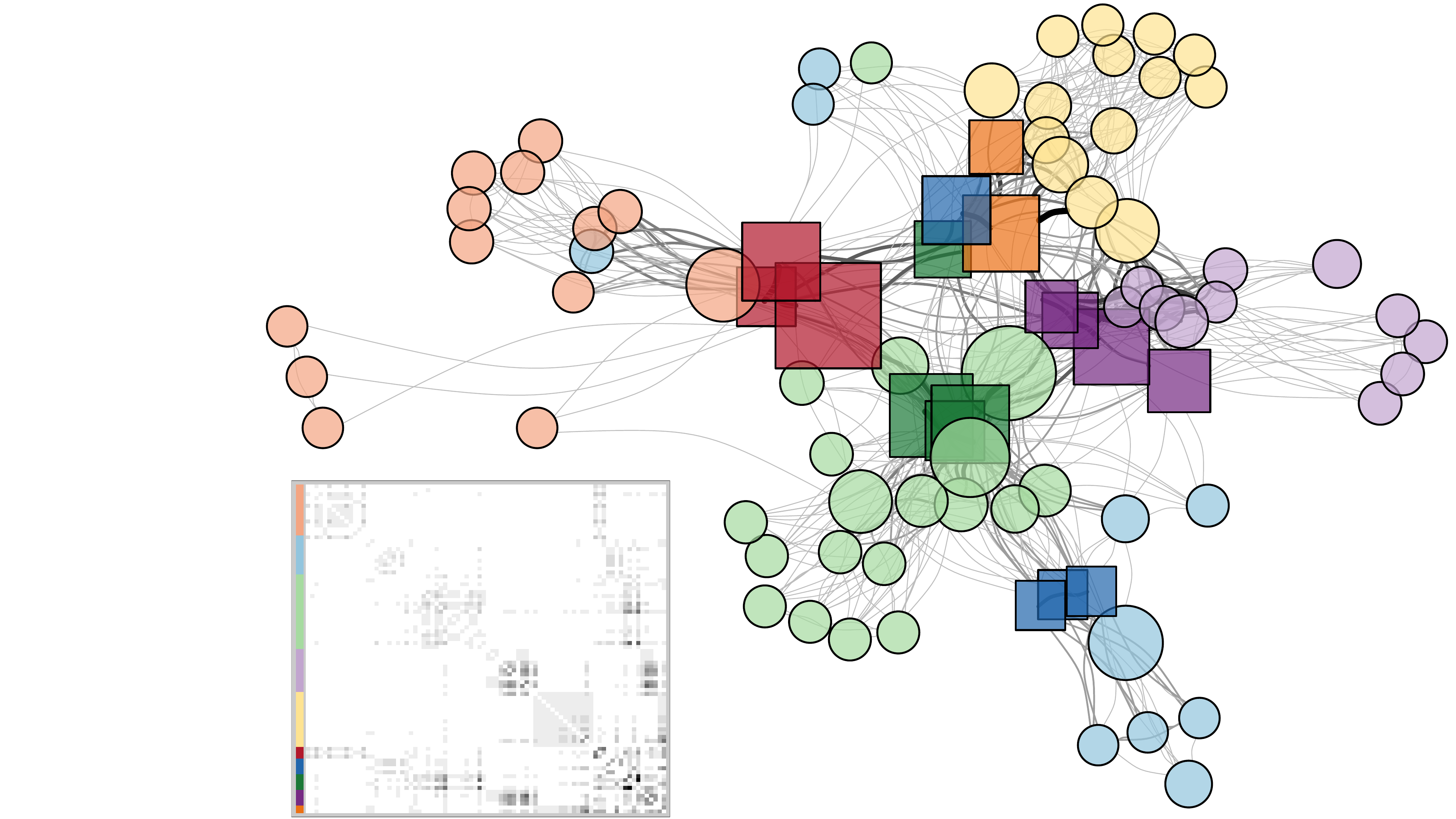}
      \vspace{0pt}
    \caption{\footnotesize{Representation of the  {\em Infinito}  network \citep[e.g.,][]{calderoni2017communities} and its adjacency matrix. Each node is a criminal and ties denote the number of co--attended summits of the 'Ndrangheta organization, as monitored during investigations. Node positions are obtained via force directed placement \citep{fru1991}, whereas  colors, both for the network and for its adjacency matrix representation, define the presumed {\em locale} membership. Darker square nodes indicate the suspected bosses of each {\em locale}, while lighter circles represent simple affiliates.  Node size is proportional to the corresponding betweenness, while tie color  and thickness is proportional to its weight. }}
    \label{figure:1}
    \vspace{-8pt}
\end{figure}

This article is motivated by an attempt to unveil unexplored knowledge on the efficiency and security structures of the 'Ndrangheta Mafia operating in the area of Milan (north of Italy) from the analysis of a network among  its members that have been monitored during a law--enforcement operation named  ``Operazione {\em Infinito}'' \citep{calderoni2017communities,legramanti2022extended}; see Figure~\ref{figure:1}. According to Interpol, 'Ndrangheta is currently one of the most widespread, proliferated and powerful criminal organizations worldwide, with a strong tendency towards transnational crime and a specific ability to generate highly structured, pervasive and resilient networks that~can penetrate not only illegal activities, but also regular business and politics \citep{paoli2007mafia,catino2014mafias,sergi2016ndrangheta}. The alarming threat and the unique challenges posed by such an organization are further~evident  from the recent establishment of the I--CAN project\footnote{\url{https://www.interpol.int/Crimes/Organized-crime/INTERPOL-Cooperation-Against-Ndrangheta-I-CAN}.}, a three year (2020--2023) initiative whose aim is to  favor across--country cooperation  and coordination in a multilateral response against 'Ndrangheta. 

Quoting the Interpol Secretary General, J\"urgen Stock: {\em ``I--CAN is about building a global early warning system against an invisible enemy''}. This description clarifies the importance of obtaining improved knowledge on the source structures of 'Ndrangheta, and crucially highlights the core challenge in addressing this goal, namely the security architecture underlying such a covert organization. As clarified within Section~\ref{sec_1}, our goal is  to cover this fundamental gap via a careful model--based approach capable of learning structure also in {\em invisible} connectivity patterns, with a focus on the analysis of the  'Ndrangheta {\em Infinito}  network retrieved from the judicial documents of ``Operazione {\em Infinito}''\footnote{Tribunale di Milano, 2011. Ordinanza di applicazione di misura coercitiva con mandato di cattura ---  art. 292 c.p.p. (Operazione Infinito). Ufficio del giudice per le indagini preliminari (in Italian)}. 

Recalling \citet{calderoni2017communities} and \citet{legramanti2022extended} ``Operazione {\em Infinito}'' has been a massive law--enforcement operation spanning six years with the aim of monitoring and disrupting the core architecture of {\em La Lombardia},  the highly--pervasive branch of the ’Ndrangheta Mafia operating in the area of Milan.~As shown in Figure~\ref{figure:1}, each node is a member of the organization, while the ties denote the number of summits (or meetings) of the criminal organization co--attended by each pair~of nodes, as monitored throughout the investigations. The original data are available at \url{https://sites.google.com/site/ucinetsoftware/datasets/covert-networks} and have been retrieved from the pre--trial detention order that was issued upon a request by the prosecution. To clarify the substantial advancements in inference on efficiency and security structures resulting~from the proposed  ZIP--SBM relative to state--of--the--art studies of such data, we consider here the pre--processed network analyzed in \citet{legramanti2022extended}, but with a crucial difference. While \citet{legramanti2022extended} focus on dichotomized ties encoding presence or absence of at least one co--attendance, we~crucially avoid such a dichotomization and leverage both sparsity and information on weighted ties to substantially expand the knowledge about  'Ndrangheta architectures and security strategies. In fact, in addition to providing a graphical evidence of grouping structures among redundant criminals, Figure~\ref{figure:1} also suggests the presence of systematic patterns both in the strength of weighted ties and in the positioning of the zero interactions. In particular, while moving from the periphery to the core, there seems to be a tendency to increasingly sparsify ties, while strengthening the non--zero ones. The dichotomization operated by  \citet{legramanti2022extended} and the community detection approach in  \citet{calderoni2017communities}  rule out these  seemingly fundamental architectures, whereas, as~clarified in Section~\ref{sec_5},  ZIP--SBM exploits such additional sources of information to extract deeper knowledge about the organization.

    \vspace{-5pt}

\begin{Remark}
\normalfont Although the newly--proposed ZIP--SBM is general and can be applied to networks measuring any form of count interaction among criminals, our focus on summit co--attendances is motivated by two main reasons. Firstly, the judicial documents of ``Operazione {\em Infinito}'' provide a highly detailed report of criminal attendances to monitored 'Ndrangheta summits. Secondly, following  \citet{calderoni2017communities} and \citet{calderoni2019nature}, the co--attendance patterns to summits are often more informative about the underlying structure and function of the networked system among the members of the criminal organization than other forms of interactions, such as, e.g., phone calls. This motivates our focus on the one--mode criminal--criminal projection of the original~two--mode criminal--summit data. Such a projection (i) is common in related studies \citep{calderoni2017communities,calderoni2019nature,cavallaro2020disrupting, ficara2021criminal, legramanti2022extended}, (ii) facilitates comparison with the previous analyses of the same one--mode {\em Infinito} network \citep[e.g.,][]{calderoni2017communities,legramanti2022extended} and  (iii) provides  access to weighted ties useful for disentangling efficiency--security structures.
\label{rem1}
\end{Remark}
    \vspace{-15pt}

\begin{Remark}
\normalfont  As a consequence of Remark~\ref{rem1}, a zero tie among two generic criminals in the network implies~that these criminals were never recorded, during the investigations, to attend a same summit. Such an event might~be either due to the actual absence of ties among these two criminals or to the fact that possible interactions have occurred through more secure mechanisms, including additional, yet secret, summits, hidden to law--enforcement investigations. The ZIP--SBM aims to disentangle these two alternatives and quantify the strength of interaction behind obscured ties. These advances are a key to  (i)  improve network reconstruction,  (ii) uncover the efficiency--security tradeoffs of the criminal organization analyzed, and  (iii)  guide the investigations towards monitoring~the inferred obscured~ties. Notice that, although the law--enforcement may fail to monitor specific summits, the~data analyzed are the result of careful investigations spanning across six years, and the  arrest warrant documents~that produced the network in Figure~\ref{figure:1} contain a comprehensive report of summit attendances. If the zero ties were only due to a general inability of law--enforcement to monitor 'Ndrangheta summits, one would expect no systematic security patterns. In fact, as discussed within  Section~\ref{sec_5}, the proposed ZIP--SBM uncovers structured obscuration mechanisms related to peculiar hierarchies within the criminal organization which are, therefore, highly unlikely to be simply the result of limited investigations.
\label{rem2}
\end{Remark}
    \vspace{-5pt}

Figure~\ref{figure:1} also suggests that the block--connectivity patterns among redundant groups might be related to the available criminal attributes encoding presumed {\em locale} affiliations and roles. Such an empirical finding is in line with forensic theories which suggest that the ’Ndrangheta organization revolves around blood family relations, often aggregated at the territorial level in structural units, known as {\em locali} \citep{paoli2007mafia,catino2014mafias,sergi2016ndrangheta}. These units administer crime in different territories and are characterized by an additional level of internal hierarchy comprising a set of affiliates and comparatively fewer bosses that oversee the illicit and licit activities within each {\em locale}, and coordinate interactions across  {\em locali}. Due to this, it is reasonable to expect that ’Ndrangheta organizations are subject to a preference toward creating redundancies within {\em locali}, rather than across such territorial units, and, at a more nested level, with respect to similarities in the role. Nonetheless, since these attributes are also a result of error--prone investigations, the inclusion of this general notion of {\em homophily} should be treated with care. As clarified in Section~\ref{sec_3}, under the proposed ZIP--SBM this is accomplished via a probabilistic reinforcement of the prior on criminal allocations to groups which favors the creation of modules that are homogenous with respect to the external attributes, but does not prevent from~inferring more nuanced group structures comprising heterogenous criminals. In fact, as illustrated in Section~\ref{sec_5}, a ZIP--SBM unveils also highly peculiar modules that depart from the  {\em locale} and role division, while pointing toward potential instabilities and obscured dynamics within the ’Ndrangheta organization.

\section{\Large 2. Bayesian Zero--Inflated Stochastic Block Models}\label{sec_2}
Let $\bY$ be the $V \times V$ symmetric adjacency matrix comprising the weighted, yet sparse, undirected ties among the $V$ criminals (nodes) in the network. More specifically, each $y_{vu}=y_{uv} \in \mathbb{N}$ measures the count relation between nodes $v$ and $u$, for every $v=2, \ldots, V$ and $u=1, \ldots, v-1$. Notice that, since our focus is on undirected networks, $y_{vu}$ is equal to $y_{uv}$ by definition. Following the discussion in Section~\ref{sec_1}, we aim to define a statistical model for $\bY$ which incorporates, and infers, (i) underlying grouping structures  defined by redundant nodes having similar connectivity patterns, (ii) flexible overt and covert block interactions among these groups, and (iii) structured sparsity to disentangle  efficiency and security architectures.

Equations \eqref{mod_tot1}--\eqref{mod_tot3} summarize the newly--proposed ZIP--SBM, whose interpretation, properties and generative construction are discussed in Sections~\ref{sec_2.1}--\ref{sec_2.2}. More specifically,  denote by $\bz=(z_1, \ldots, z_V)$  the vector of criminal allocations to redundant groups, where $z_v=h$ indicates that the $v$--th criminal belongs to group $h \in \{1, \ldots, H\}$, for each $v=1, \ldots, V$. Moreover, let $\bar{\bPi}$ and~$\bar{\bLambda}$ be two $H \times H$ symmetric matrices whose entries $\bar{\pi}_{hk} \in (0,1)$ and $\bar{\lambda}_{hk} \in \mathbb{R}^+$ correspond to the zero--inflation probability (security) and the rate (efficiency), respectively,~of~the~ties among generic criminals in groups $h$ and $k$, for each $h=1, \ldots, H$ and $k=1, \ldots, h$. Then, the proposed Bayesian  ZIP--SBM is defined as
\begin{eqnarray}
&&(y_{vu} \mid z_v=h,z_{u}=k, \bar{\pi}_{hk}, \bar{\lambda}_{hk}) \sim \textsc{zip}(\bar{\pi}_{hk}, \bar{\lambda}_{hk}), \quad \mbox{independently for} \ v=2, \ldots, V, \ \ u=1, \ldots, v-1,  \quad \ \label{mod_tot1} \\
&&\bar{\pi}_{hk} \sim \mbox{Beta}(a,b), \quad \bar{\lambda}_{hk} \sim \mbox{Gamma}(a_1,a_2),  \qquad  \  \ \ \ \mbox{independently for} \ h=1, \ldots, H, \ \ k=1, \ldots, h,  \ \label{mod_tot2} \\
&&(\bz =(z_1, \ldots, z_V)  \mid \bc)\sim \textsc{gn}(\gamma; \bc),  \ \label{mod_tot3}
\end{eqnarray}
where $\textsc{zip}(\bar{\pi}_{hk}, \bar{\lambda}_{hk})$ in \eqref{mod_tot1} denotes the zero--inflated Poisson distribution  \citep[e.g.,][]{lambert1992zero,ghosh2006bayesian} with probability mass function $p(y)=\bar{\pi}_{hk} \mathbbm{1}(y=0)+ (1-\bar{\pi}_{hk})\bar{\lambda}_{hk}^{y}\mbox{e}^{-\bar{\lambda}_{hk}}/y!$, $(y  \in \mathbb{N})$, for the ties among pairs of criminals in groups $h$ and $k$, respectively, whereas equations \eqref{mod_tot2}--\eqref{mod_tot3} clarify the selected prior distributions for the block--specific parameters in $\bar{\bPi}$ and $\bar{\bLambda}$, and  for the allocation vector $\bz$ encoding membership of criminals to the underlying groups. In particular, in equation~\eqref{mod_tot2} we follow the recommended practice in Bayesian stochastic block models for binary \citep[e.g.,][]{nowicki2001estimation,geng2019probabilistic} and weighted \citep[e.g.,][]{mcdaid2013improved} networks, and assume independent $\mbox{Beta}(a,b)$ and  $\mbox{Gamma}(a_1,a_2)$ priors for the entries $\bar{\pi}_{hk}$~in  $\bar{\bPi}$, and $\bar{\lambda}_{hk}$ in $\bar{\bLambda}$,~respectively. As discussed in Section~\ref{sec_2.1} this choice guarantees conjugacy properties that facilitate posterior inference. In~addition, it introduces suitable dependence structures among the ties, both overt and covert, between criminals  in the network. For the allocation vector we adopt in~\eqref{mod_tot3} a supervised version $\textsc{gn}(\gamma; \bc)$  of the Gnedin process prior \citep{Gnedin2010} which belongs to the general Gibbs--type class \citep[e.g.,][]{de_2013} employed in   \citet{legramanti2022extended}. As clarified in Section~\ref{sec_2.2}, this prior naturally characterizes the affiliation mechanism of criminals to the redundant groups via a sequential scheme, which depends both on the network size and also on the current number, dimension and, possibly, the attribute composition $\bc$ of such groups. In the study of the {\em Infinito} network in~Figure~\ref{figure:1}, the external attribute vector $\bc=(c_1, \ldots, c_V) \in \{1, \ldots, C\}^V$~encodes memberships of criminals to a known exogenous partition that corresponds to a combination of {\em locale}--role information and, hence, is expected to influence the formation of the redundant endogenous groups encoded in $\bz$. Notice that,~as discussed in Section~\ref{sec_2.2},  when $\bc$ is not available, the ZIP--SBM can be still implemented by replacing the~supervised Gnedin process prior $\textsc{gn}(\gamma; \bc)$ with its unsupervised counterpart $\textsc{gn}(\gamma)$.

As outlined within Section~\ref{sec_1.0}, the ZIP--SBM model in \eqref{mod_tot1}--\eqref{mod_tot3} generalizes the extended stochastic block model of  \citet{legramanti2022extended}, originally developed for binary networks, in order to include weighted ties while accounting for possible zero--inflation mechanisms pointing to systematic sparsity patterns that may inform on security strategies. This is accomplished by replacing the commonly--assumed Bernoulli distribution within each block with a zero--inflated Poisson. Albeit providing a natural methodological extension,  this choice substantially expands the potential for improved inference within the context of criminal networks and opens new avenues to infer yet--unexplored generative architectures produced by efficiency--security tradeoffs \citep{morselli2007efficiency,catino2015mafia}. In fact, as discussed within Section~\ref{sec_1.0}, unlike other stochastic block models incorporating zero--inflation  \citep{aicher2015learning,mariadassou2015convergence,ng2021weighted}, the proposed ZIP--SBM does not require  knowledge about the nature of zero ties, but rather aims to uncover it as part of the inference process; see Section~\ref{sec_2.1} for additional details, and refer to the empirical studies in Sections~\ref{sec_4}--\ref{sec_5} for a clear illustration of the applied potential of the ZIP--SBM in \eqref{mod_tot1}--\eqref{mod_tot3} relative to state--of--the--art alternatives.

\vspace{8pt}
\subsection{\large 2.1. Generative construction, interpretation and properties}\label{sec_2.1}
\vspace{2pt}
Adapting classical augmented--data representations of zero--inflated Poisson distributions   \citep[e.g.,][]{lambert1992zero,ghosh2006bayesian} to our network setting, model \eqref{mod_tot1} can be readily obtained by marginalizing out the latent variables $w_{vu}\in \mathbb{N}$ and $x_{vu} \in \{0;1\}$ in the  following generative representation
\begin{equation}
\begin{split}
&y_{vu}=w_{vu}(1-x_{vu}), \\
&(w_{vu}\mid z_v=h,z_{u}=k, \bar{\lambda}_{hk}) \sim \mbox{Poisson}(\bar{\lambda}_{hk}), \quad (x_{vu} \mid z_v=h,z_{u}=k, \bar{\pi}_{hk}) \sim \mbox{Bern}(\bar{\pi}_{hk}),
\label{eq1}
\end{split}
\end{equation}
independently for  $v=2, \ldots, V$ and $u=1, \ldots, v-1$.  Combining \eqref{eq1} with \eqref{mod_tot2}--\eqref{mod_tot3} clarifies that the proposed ZIP--SBM can be alternatively reinterpreted as the combination of two underlying stochastic block models for the augmented networks $\bW$ and $\bX$ with entries $w_{vu}$ and $x_{vu}$, respectively, for each $v=2, \ldots, V$ and $u=1, \ldots, v-1$.   The former encodes the observed or latent weighted ties among each generic pair of criminals $(v,u)$, for every $v=2, \ldots, V$ and $u=1, \ldots, v-1$, whereas the latter characterizes the excess of zeros relative to those expected under the distribution of such weighted ties, thus informing on possible security architectures.

\begin{figure}[t!]
\centering
\vspace{-5pt}
    \includegraphics[trim=3.5cm 4cm 1cm 0.5cm,clip,width=1\textwidth]{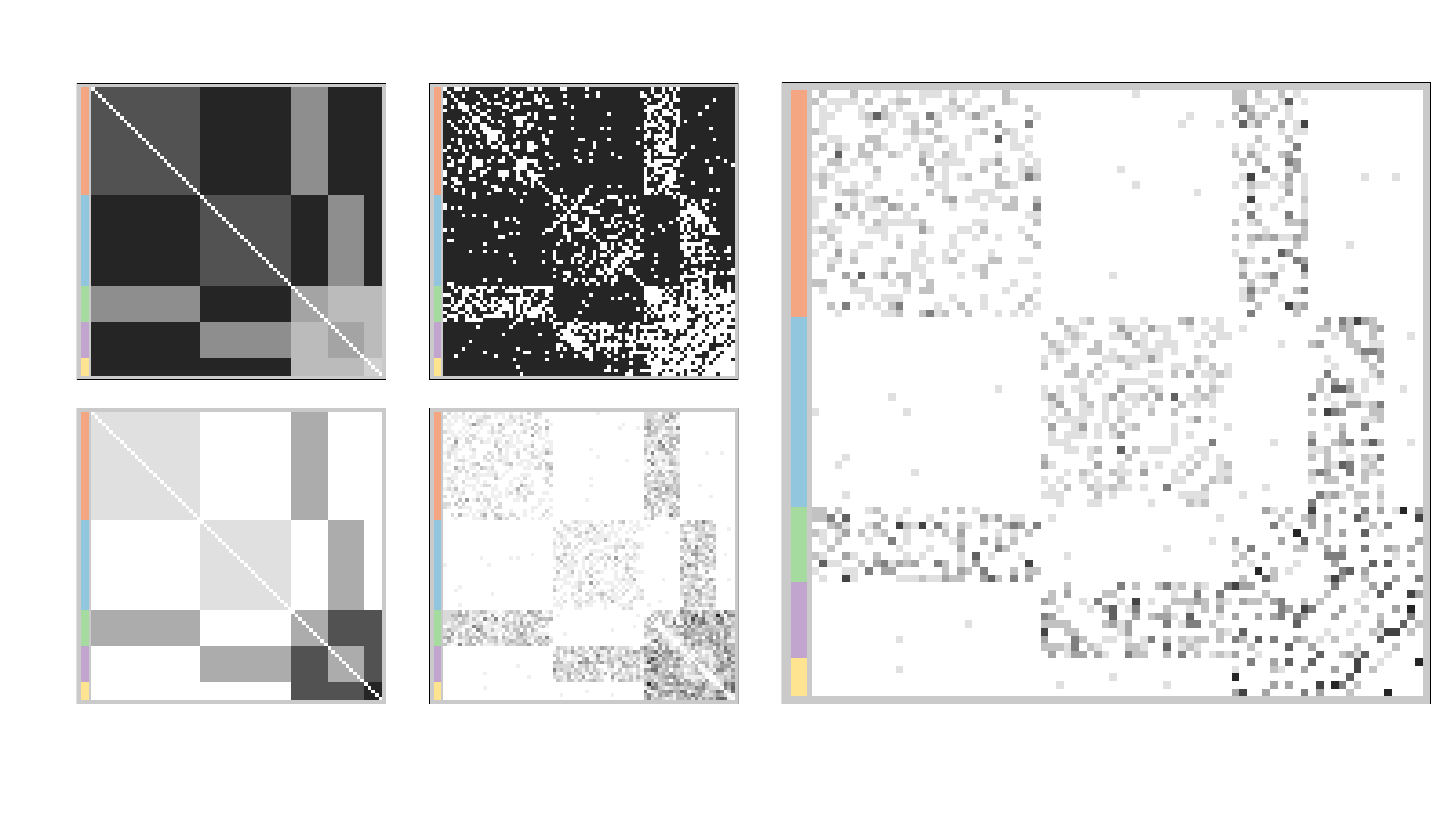}
    \put(-433,0){${\bf \Lambda}$}
        \put(-310,0){${\bf W}$}
            \put(-445,238){${\bf 1}-{\bPi}$}
        \put(-317,238){${\bf 1}-{\bf X}$}
                \put(-150,238){${\bf Y}={\bf W} \odot ({\bf 1}-{\bf X})$}
    \caption{\footnotesize{Graphical representation of the generative mechanism underlying ZIP--SBM. Side colors correspond to the  grouping structure encoded in $\bz$. Criminals allocated to the same group are redundant from a connectivity perspective and, hence, the associated rows in ${\bf 1}-\bPi={\bf 1}-\bZ\bar{\bPi}\bZ^{\intercal}$ and $\bLambda=\bZ\bar{\bLambda}\bZ^{\intercal}$ are equal by construction. Notice that  $\bZ$ is the $V \times H$ matrix with rows $\bz_v=[\mathbbm{1}(z_v=1), \ldots,  \mathbbm{1}(z_v=H)]$ for $v=1, \ldots, V$. The matrices ${\bf 1}-\bX$ and $\bW$ have generic entries $1-x_{vu}$ and $w_{vu}$, respectively, simulated as in~\eqref{eq1}, where $1-\pi_{vu}=1-\bz_v\bar{\bPi}\bz_u^{\intercal}$ and $\lambda_{vu}=\bz_v\bar{\bLambda}\bz_u^{\intercal}$ correspond to the entries in position $(v,u)$ of ${\bf 1}-\bPi$ and $\bLambda$, respectively.   Recalling \eqref{eq1},  the element--wise Hadamard product $\odot$ between $\bW$ and ${\bf 1}-{\bf X}$  yields the observed network $\bY$.
  \vspace{2pt}}}
    \label{figure:2}
\end{figure}

As is clear from representation~\eqref{eq1}, the distribution of both quantities $w_{vu}$ and $x_{vu}$ only depends on the group allocations $z_v$ and $z_u$ of the two involved nodes $v$ and $u$, and on the block--specific parameters defining the ties among such groups. As a consequence, these parameters are shared~among all ties involving criminals from the same pair of groups $(h,k)$, and only change as a function of~$(h,k)$.~This construction crucially leverages, formalizes, and infers redundancy structures made by groups of criminals with a similar position within the network topology and, hence, redundant from a connectivity perspective  \citep[e.g.,][]{Bouchard,cavallaro2020disrupting}. To clarify this point, notice that as a consequence of  \eqref{eq1}, the individual parameters $\lambda_{vu}$ and $\pi_{vu}$ indexing the distribution of $w_{vu}$ and $x_{vu}$, respectively, for each  $v=2, \ldots, V$ and $u=1, \ldots, v-1$, can be derived as 
\begin{equation}
\lambda_{vu}=\bar{\lambda}_{z_v,z_u}=\bz_v\bar{\bLambda}\bz_u^{\intercal} \quad \mbox{and} \quad \pi_{vu}=\bar{\pi}_{z_v,z_u}=\bz_v\bar{\bPi}\bz_u^{\intercal},  \qquad \mbox{for} \ v=2, \ldots, V, \ u=1, \ldots, v-1,
\label{eq5}
\end{equation}
where $\bz_v=[\mathbbm{1}(z_v=1), \ldots,  \mathbbm{1}(z_v=H)]$ and $\bz_u=[\mathbbm{1}(z_u=1), \ldots,  \mathbbm{1}(z_u=H)]$ comprise vectors of all zero entries except for a single $1$ in the position corresponding to the group indicator to which the nodes $v$ and $u$  have been allocated, respectively. Let $\bLambda$ and $\bPi$ denote the $V \times V$ symmetric matrices with entries  $\lambda_{vu}$ and $\pi_{vu}$, respectively, for each  $v=2, \ldots, V$ and $u=1, \ldots, v-1$. Then the above result implies that the rows of $\bLambda$ and $\bPi$  corresponding to criminals allocated to the same redundant group are equal, respectively, meaning that any two criminals  $v$ and $v'$ within the same group share the same rates of interaction and zero--inflation probabilities in the ties with others, i.e.,  $\lambda_{vu}=\lambda_{v'u}$ and $\pi_{vu}=\pi_{v'u}$ for every $u$ different from $v$ and $v'$. This enforces, in turn, equality in distribution among the associated rows of $\bW$ and $\bX$, respectively, thereby incorporating the precise notion of stochastic equivalence \citep[e.g.,][]{nowicki2001estimation}. Due to the definition of $y_{vu}$ in \eqref{eq1}, the same holds for the observed network $\bY$. See Figure~\ref{figure:2} for a graphical illustration of the generative model underlying $\bY$, which further stresses these concepts. Such a figure also clarifies that the overall sparsity of the network is controlled by both the zero--inflation probabilities in $\bPi$ and the Poisson rates in $\bLambda$. High values of the entries in $\bPi$  and/or low rates in $\bLambda$ yield sparser networks. As clarified in the simulation studies within Section~\ref{sec_4}, although the parameters associated with sparser blocks are more difficult to learn due to the presence of relatively--few weighted ties,  the proposed ZIP--SBM generally achieves accurate performance also in these settings.

Besides providing a simple augmented--data representation for the formation process of the ties in $\bY$ which~is useful for both estimation and inference, the above formulation characterizes also a natural and interpretable generative model for criminal networks. In particular, according to~representation \eqref{eq1} and Figure~\ref{figure:2}, the observed tie $y_{vu} \in \mathbb{N}$ among criminals $v$ and $u$, depends on the decision to either obscure or not ties among such criminals, encoded in $x_{vu} \in \{0;1\}$, and on the corresponding, observed or latent, strength of interaction measured by~$w_{vu} \in \mathbb{N}$. Following this line of reasoning, if $y_{vu}>0$ then $(x_{vu}=0,w_{vu}=y_{vu})$, meaning that no security strategy has been adopted for the pair $(v,u)$ and the weighted tie among $v$ and $u$ has been actually observed. A zero count $y_{vu}=0$ can be instead associated with two different scenarios, namely $(x_{vu}=1, w_{vu} \in  \mathbb{N})$ or $(x_{vu}=0,w_{vu}=0)$. In the first case, ties among $v$ and $u$ have been arguably obscured --- no matter whether the strength of these ties is high or low. In the second situation, no security strategy has been adopted, but the actual count of interactions among $v$ and $u$ is effectively zero. Recalling  \eqref{eq1}, these two cases can be disentangled under the proposed ZIP--SBM via the conditional probability
\begin{equation}
\begin{split}
\mbox{pr}(x_{vu}=1 \mid y_{vu}=0,z_v=h,z_{u}=k, \bar{\pi}_{hk}, \bar{\lambda}_{hk})&=\frac{\mbox{pr}(x_{vu}=1, y_{vu}=0 \mid z_v=h,z_{u}=k, \bar{\pi}_{hk})}{\mbox{pr}(y_{vu}=0 \mid z_v=h,z_{u}=k, \bar{\pi}_{hk}, \bar{\lambda}_{hk})}, \\
 &=\frac{\bar{\pi}_{hk}}{\bar{\pi}_{hk}+(1-\bar{\pi}_{hk})\mbox{e}^{-\bar{\lambda}_{hk}}},
\label{eq3}
\end{split}
\end{equation} 
for every $v=2, \ldots, V$ and $u=1, \ldots, v-1$. When $y_{vu}=0$, it is also of  interest to assess to what extent such a zero tie corresponds to an underlying non--zero interaction. Recalling our previous~discussion, when a security strategy is implemented,~i.e., $x_{vu}=1$, then $y_{vu}=0$ no matter whether an actual underlying tie $w_{vu}$ is effectively present, i.e., $w_{vu}>0$, or not, namely $w_{vu}=0$. Disentangling these two alternatives is a key in law enforcement to unveil the effective structure of the underlying criminal network and assess which obscured ties are hiding strong interactions that are worth investigations. Information on non--zero obscured ties can be obtained under the proposed model via the conditional probability 
\begin{equation}
\begin{split}
\mbox{pr}(w_{vu}>0  \mid y_{vu}=0,z_v=h,z_{u}=k, \bar{\pi}_{hk}, \bar{\lambda}_{hk})&=\frac{\mbox{pr}(w_{vu}>0, y_{vu}=0 \mid z_v=h,z_{u}=k, \bar{\pi}_{hk},\bar{\lambda}_{hk})}{\mbox{pr}(y_{vu}=0 \mid z_v=h,z_{u}=k, \bar{\pi}_{hk}, \bar{\lambda}_{hk})}, \\
 &=\frac{(1-\mbox{e}^{-\bar{\lambda}_{hk}})\bar{\pi}_{hk}}{\bar{\pi}_{hk}+(1-\bar{\pi}_{hk})\mbox{e}^{-\bar{\lambda}_{hk}}},
\label{eq31}
\end{split}
\end{equation} 
for every $v=2, \ldots, V$ and $u=1, \ldots, v-1$. Note that, a very large $\bar{\lambda}_{hk}$ implies $e^{-\bar{\lambda}_{hk}} \approx 0$, and,~therefore, both~\eqref{eq3} and \eqref{eq31} are almost equal and coincide approximately with $ \bar{\pi}_{hk}/\bar{\pi}_{hk} =1$. This  provides a reasonable result stating that, if $\bar{\lambda}_{hk}$ is very large, any observed zero tie is almost surely due to a secrecy strategy (i.e.,~$\mbox{pr}(x_{vu}=1 \mid y_{vu}=0,z_v=h,z_{u}=k, \bar{\pi}_{hk}, \bar{\lambda}_{hk})\approx 1$) hiding a non--zero interaction ($\mbox{pr}(w_{vu}>0  \mid y_{vu}=0,z_v=h,z_{u}=k, \bar{\pi}_{hk}, \bar{\lambda}_{hk}) \approx 1$). This is because  the probability of a zero tie under a Poisson with a very large rate is essentially $0$ and, hence, any observed zero tie is almost surely the result of a secrecy strategy arising from the zero--inflation mechanism. In this context, almost all the zero ties can be used to effectively infer $ \bar{\pi}_{hk}$.

Removing the conditioning on $y_{vu}=0$ in \eqref{eq31}, provides instead a~measure of efficiency which quantifies the probability that  $v$ and $u$ establish a  non--zero tie, irrespectively of whether it has been obscured or not. This~yields
\begin{equation}
\mbox{pr}(w_{vu}>0 \mid z_v=h,z_{u}=k,\bar{\lambda}_{hk})=1-\mbox{e}^{-\bar{\lambda}_{hk}},
\label{eq4}
\end{equation} 
for each $v=2, \ldots, V$ and $u=1, \ldots, v-1$.

As detailed in \eqref{eq3}, evidence of security structures behind a zero tie $y_{vu}=0$ among two generic criminals $v$~and $u$ allocated to groups $h$ and $k$, respectively,  does not necessarily require a high~$ \bar{\pi}_{hk}$. In fact, even~a~low~$ \bar{\pi}_{hk}$~can point toward an {\em unusual} zero tie, when  $\bar{\lambda}_{hk}$ is large, since such a zero will appear as highly unlikely under the  $ \mbox{Poisson}(\bar{\lambda}_{hk})$ distribution for the weighted ties. As a consequence, it is fundamental to accurately infer $ \bar{\pi}_{hk}$  and  $\bar{\lambda}_{hk}$,  for each  $h=1, \ldots, H$ and $k=1, \ldots, h$, from the observed data $y_{vu}$, $v=2, \ldots, V$ and $u=1, \ldots, v-1$ in $\bY$. This is also a key to quantify evidence of efficiency via \eqref{eq4}. As clarified in \eqref{mod_tot1}--\eqref{mod_tot3}, the proposed ZIP--SBM achieves this goal by expressing the network $\bY$ as a function of criminal allocations to underlying groups and a lower number of block--specific parameters encoding efficiency--security patterns within and across these groups. This means that all the ties connecting criminals in group $h$ with those in group~$k$ are conditionally independent realizations from a common zero--inflated Poisson distribution with parameters $(\bar{\pi}_{hk},\bar{\lambda}_{hk})$. Therefore,~$(\bar{\pi}_{hk},\bar{\lambda}_{hk})$ can  be effectively inferred, for every $h=1, \ldots, H$ and $k=1, \ldots, h$, leveraging the information from all the~ties among criminals allocated to groups $h$ and $k$, respectively. Recalling, for example, \citet{li2012} both parameters are identifiable under the zero--inflated Poisson we consider within each block.

As mentioned before, the inclusion of group structures encoded in $\bz$, not only facilitates inference on efficiency--security tradeoffs measured by the zero--inflated Poisson parameters, but also allows one to obtain evidence of redundancy patterns in the criminal network, along with the corresponding sizes and composition. As a result, the proposed formulation crucially accounts for the two fundamental sources of resilience in criminal organizations, namely efficiency--security via $(\bar{\bPi},\bar{\bLambda})$  and redundancy through $\bz$. In model \eqref{mod_tot1}--\eqref{mod_tot3}, these parameters are inferred from a Bayesian perspective. Within a criminal network context characterized by error--prone measurements and expert knowledge from criminology, such a perspective is particularly suitable to facilitate principled uncertainty quantification and formal inclusion of prior information. 

The inclusion of prior distributions further allows one to introduce dependence among the covert and overt ties in the network, which is expected in the context of criminal networks. In~fact, although \eqref{eq1} implies that  variables $x_{vu}$ and $w_{vu}$ are conditionally independent, for each $v=2, \ldots, V$ and $u=1, \ldots, v-1$, given the allocation vector and the block--specific parameters, by marginalizing out these latter quantities from the joint likelihood of $\bX$ and $\bW$ introduces dependence among the entries of these two matrices. More specifically, exploiting conjugacy induced by \eqref{mod_tot2} together with representation \eqref{eq1} of model  \eqref{mod_tot1}, it is possible to marginalize out analytically $\bar{\bPi}$ and $\bar{\bLambda}$ in $p(\bX \mid \bz,\bar{\bPi})$ and $p(\bW \mid \bz,\bar{\bLambda})$, respectively, to obtain
\begin{equation}
\begin{split}
p(\bX \mid \bz)&=\prod\nolimits_{h=1}^{H} \prod\nolimits_{k=1}^h \frac{\mbox{B}(a+x_{hk}, b+ n_{hk}-x_{hk})}{\mbox{B}(a,b)},\\
p(\bW \mid \bz)&=\left(\prod\nolimits_{v=2}^V\prod\nolimits_{u=1}^{v}w_{vu}!\right)^{-1}\prod\nolimits_{h=1}^{H} \prod\nolimits_{k=1}^h\frac{a_2^{a_1} \Gamma(a_1+w_{hk})}{(a_2+n_{hk})^{a_1+w_{hk}}\Gamma(a_1)},
\end{split}
 \label{eq10}
\end{equation}
where $\mbox{B}(\cdot,\cdot)$ and $\Gamma(\cdot)$ are the Beta and Gamma functions, $n_{hk}$ is the total number of unique pairs involving a criminal in group $h$ and one in group $k$, whereas $x_{hk}$ and $w_{hk}$ denote the sum, over such pairs, of the entries $x_{vu}$ and $w_{vu}$ in $\bX$ and $\bW$, respectively. Let $\bX_{-(v,u)}$ and  $\bW_{-(v,u)}$ be the matrices $\bX$ and  $\bW$ excluding the entries $x_{vu}$ and $w_{vu}$, respectively. Then, if $z_v=h$ and $z_u=k$, direct application of  \eqref{eq10} yields the  predictive probabilities 
\begin{equation}
\begin{split}
\mbox{pr}(x_{vu}=1 \mid \bX_{-(v,u)}, \bz)&=\frac{p(x_{vu}=1, \bX_{-(v,u)} \mid \bz)}{p(\bX_{-(v,u)} \mid \bz)}\\
&=\frac{\mbox{B}(a+x^{-(v,u)}_{hk}+1, b+ n_{hk}^{-(v,u)}-x^{-(v,u)}_{hk})}{\mbox{B}(a+x^{-(v,u)}_{hk}, b+ n_{hk}^{-(v,u)}-x^{-(v,u)}_{hk})}=\frac{a+x^{-(v,u)}_{hk}}{a+b+n^{-(v,u)}_{hk}},\\
&\\
\mbox{pr}(w_{vu}=w \mid \bW_{-(v,u)}, \bz)&=\frac{p(w_{vu}=w, \bW_{-(v,u)} \mid \bz)}{p(\bW_{-(v,u)} \mid \bz)}\\
&=\frac{1}{w!}\frac{ \Gamma(a_1+w^{-(v,u)}_{hk}+w)}{(a_2+n^{-(v,u)}_{hk}+1)^{a_1+w^{-(v,u)}_{hk}+w}}\frac{(a_2+n^{-(v,u)}_{hk})^{a_1+w^{-(v,u)}_{hk}}}{\Gamma(a_1+w^{-(v,u)}_{hk})},
\end{split}
 \label{eq10.p}
\end{equation}
for each $v=2, \ldots, V$ and $u=1, \ldots, v-1$, where the apex $^{-(v,u)}$ denotes all the previously--defined quantities, evaluated disregarding the pair of nodes $(v,u)$. Equation~\eqref{eq10.p} clarifies that the conditional distribution of each covert and overt tie between $v$ and $u$ is influenced by those among nodes allocated to the same pair of groups, as expected in redundant modules within criminal networks. Notice that such a form of dependence is even more general than the classical Markov one adopted in popular exponential random graph models \citep{frank1986markov}, since it allows two ties to be dependent also if there is no node in common, as long as these nodes are allocated to the same pair of groups. Dependence across blocks and between $\bX$ and $\bW$ is instead induced by the shared partition $\bz$ and its prior in \eqref{mod_tot3}, which in turn translates into dependencies among the ties within~the  observed network $\bY$. See Section~\ref{sec_2.2} below for a detailed presentation and for a constructive motivation of the assumed supervised Gnedin process prior for $\bz$.

\vspace{8pt}

\subsection{\large 2.2. Redundancy via supervised Gnedin process prior}\label{sec_2.2}
\vspace{2pt}

The definition of a prior for the group membership vector $\bz$ is a challenging task since it requires a carefully--tailored distribution for a random partition which can incorporate realistic mechanisms of criminals affiliation to redundant modules, while enabling uncertainty quantification on the unknown number of underlying groups and inclusion of information from exogenous criminal attributes $\bc$, when available.

Focusing first on the simpler case in which exogenous attributes $\bc$ are not available, routine implementations of stochastic block models rely on $\mbox{Dirichlet--multinomial}(\alpha, \overline{H})$ priors for $\bz$ \citep[][]{nowicki2001estimation}, where  $\overline{H} \geq H$ denotes the total number of groups in the whole population  of criminals, including also yet unobserved clusters occupied by criminals that have not been investigated. These~Dirichlet--multinomial priors are obtained by marginalizing out in  $(z_{v} \mid \btheta)\sim \mbox{Multinomial}(1, \btheta=(\theta_1, \ldots, \theta_{\overline{H}}))$, independently for $v=1, \ldots, V$, the vector of group membership probabilities  $\btheta$ distributed according to the $\mbox{Dirichlet}(\balpha=(\alpha, \ldots, \alpha))$. Unfortunately, this choice is practically and~conceptually suboptimal within our context since it requires one to pre--select $\overline{H}$, which is clearly not known~in practice.  Notice that $\overline{H}$ should not be confused with $H$ which denotes, instead, the total number of groups occupied by the $V$ observed criminals under analysis. In fact, $H \leq \mbox{min}\{V, \overline{H}\}$.
An alternative to pre--specifying~$\overline{H}$ would be to assume that $\overline{H} \rightarrow \infty$ when $V \rightarrow \infty$, as done in the infinite relational model \citep{kemp_2006,schmidt_2013}. However, this setting is not realistic in networks associated with organized crime where, as a consequence of predefined rules and pyramidal structures \citep{paoli2007mafia,catino2014mafias,catino2015mafia}, it is more reasonable to expect that the total number of groups  $\overline{H}$ is finite, rather than infinite, even~in~the whole, possibly infinite,~population of criminals. 

To overcome the above challenges, we employ a mixture--of--finite--mixtures construction \citep[see e.g.,][]{geng2019probabilistic,legramanti2022extended} which leverages the Dirichlet--multinomial prior for~$\bz$, but crucially lets  $\overline{H}$ to be finite and random, rather than finite and fixed as in the more classical specifications. In particular, for~$a>0$, let $(a)_n=a(a+1) \cdots (a+n-1)$ with $(a)_0=1$. Then, a sensible and computationally--tractable choice in this context is to consider the Gnedin process prior \citep{Gnedin2010,de_2013}, which is obtained by marginalizing out in $(\bz \mid \overline{H}) \sim \mbox{Dirichlet--multinomial}(\alpha=1, \overline{H})$, the random number of groups $\overline{H}$ with probability mass function $p(\overline{H})=\gamma(1-\gamma)_{\overline{H}-1}/\bar{H}!$  for $\overline{H} \in \{1, 2, \ldots \}$, where $(1-\gamma)_{\overline{H}-1}=(1-\gamma)(2-\gamma)(3-\gamma)\cdots(\overline{H}-1-\gamma)$, whereas $\gamma \in (0,1)$ is the prior  hyper--parameter  discussed in detail below. Notice that such a probability mass function for \smash{$\overline{H}$} has heavy tails~and mode at $1$, thus favoring parsimonious  reconstructions of redundancy patterns in criminal networks, while maintaining a level of flexibility to account for more complex modular architectures composed by a large number of  groups. Moreover, such a prior belongs to the general Gibbs--type class explored by  \citet{legramanti2022extended} within the context of stochastic block models for binary networks, and, unlike the one considered in \citet{geng2019probabilistic}, crucially admits an urn--scheme representation which facilitates interpretation and inference. In particular, let $n_{h,-v} $ and $H_{-v}$ denote the cardinality of group $h$ and the total number of non--empty groups, respectively, after removing the $v$--th criminal. Then, leveraging such an urn scheme, the prior distribution  over the group assignments for criminal $v$, conditioned on the memberships $\bz_{-v}=(z_1, \ldots, z_{v-1}, z_{v+1}, \ldots, z_V)$ of the other $V-1$  is defined as $\mbox{pr}(z_v=h \mid \bz_{-v})  \propto (n_{h,-v}+1)[(V-1)-H_{-v}+\gamma]$ if $h$ is an already--occupied group by the other criminals, excluding the $v$--th one, i.e., $h=1, \ldots, H_{-v}$. Conversely, if $h$ corresponds to a new group, i.e., $ h=H_{-v}+1$, then  $\mbox{pr}(z_v=h \mid \bz_{-v})  \propto H_{-v}(H_{-v}-\gamma)$. Such a representation also clarifies the role of the tuning hyper--parameter $\gamma \in (0,1)$ in  controlling the formation of yet--unseen groups; the higher $\gamma$ is, the lower  the probability of generating new groups. As highlighted in Sections~\ref{sec_4} and \ref{sec_5}, our empirical results are robust to the choice of $\gamma$, and its impact on inference is clearly milder than pre--assuming that $\overline{H}$ is fixed~and~equal~to~a~single, yet unknown, value or  infinite.

According to the above urn scheme, when $v$ is a generic criminal entering the network among the other~$V-1$ members, such a criminal can either join an already--occupied group of redundant members and inherit the associated efficiency and security architectures, or can create a yet--unseen group with possibly different patterns in the zero--inflation probabilities and rates of interactions with others.  This affiliation process crucially depends on the size of the criminal network, the current number  and cardinality of non--empty groups and, finally, the tuning parameter $\gamma \in (0,1)$. All these dimensions are at the core of the structured recruiting process~and~growth~in complexity of criminal organizations \citep[see, e.g.,][]{catino2014mafias,catino2015mafia}, thereby yielding a realistic construction. For example, the presence of $n_{h,-v}$ in the above urn scheme allows the inclusion of a {\em rich get richer} property which is realistic for those groups of highly operative criminals that are more visible to law enforcement, thus requiring an increased redundancy to preserve resilience.

Although the classical Gnedin process prior provides a sensible construction, it does not account for information on those exogenous criminal attributes that are often available in law--enforcement investigations. For example, following Section~\ref{sec_1.1}, in the {\em Infinito} network we possess additional information on presumed {\em locale} membership and role for each criminal. Even if it is unrealistic to believe that the redundancy patterns encoded in $\bz$ perfectly overlap with the  exogenous partition of criminals provided by these possibly error--prone attributes, excluding this additional information in the prior for $\bz$ is similarly--suboptimal. In fact, following criminology theories on trust, human capital and rules 
\citep[e.g.,][]{campana2013cooperation,charette2017network,Bouchard,cavallaro2020disrupting,berlusconi2022come} it is reasonable to expect that a given criminal is more likely to join groups mainly composed by members with the same combination of {\em locale}--role, rather than the opposite. To incorporate this information, we tailor the product partition model  in \citet{legramanti2022extended} to the Gnedin process prior for obtaining the supervised version $\textsc{gn}(\gamma; \bc)$ in \eqref{mod_tot3}  which favors the formation of groups that are homogenous with respect to the attributes of the criminals. More specifically, let $\bc=(c_1, \ldots, c_V) \in \{1, \ldots, C\}^V$ encode the memberships of criminals to a known exogenous partition which, within the {\em Infinito} network study, corresponds to a combination of {\em locale}--role information considered also in \citet{legramanti2022extended}.  Then, the assumed $\textsc{gn}(\gamma; \bc)$ prior in \eqref{mod_tot3} incorporates $\bc$  via the following urn scheme
\begin{equation}
 \label{eq8}
\begin{aligned}
& \mbox{pr}(z_v=h \mid \bz_{-v},\bc) \propto \begin{cases} \scalemath{1}{\frac{n_{hc_v, -v}+\alpha_{c_v}}{n_{h,-v}+\alpha_0}} (n_{h,-v}+1)[(V-1)-H_{-v}+\gamma] &  \text{for} \  h=1, \ldots, H_{-v}, \\ \scalemath{1}{\frac{\alpha_{c_v}}{\alpha_0}} [H_{-v}(H_{-v}-\gamma)]&  \text{for} \  h=H_{-v} + 1, \\ \end{cases} 
\end{aligned}
\end{equation}
where $\alpha_1, \ldots, \alpha_C$ are positive cohesion parameters whose sum is $\alpha_0$, whereas $ n_{h c_v, -v}$ denotes the number~of criminals within group $ h $ that belong to the same exogenous partition of the $v$--th one. In the context of the {\em Infinito} network study, this means that the group--allocation probabilities of the classical, unsupervised, Gnedin process prior are now inflated or deflated in \eqref{eq8} by a term which induces a probabilistic homophily  favoring the allocation of criminal $v$ to those groups containing a higher fraction of existing members with the same {\em locale}--role  combination. In~the motivating application, the exogenous class of each simple affiliate corresponds to the associated {\em locale}, whereas all bosses have a unique label indicating that such members cover a leadership position. Finally, a subgroup of affiliates belonging to the purple {\em locale} who are known from the judicial documents to cover a peripheral role are given a distinct label.  Being probabilistic, such a reinforcement does not preclude the formation of more heterogenous groups, when necessary; see Section~\ref{sec_5}. As clarified in Section~\ref{sec_3},~this~prior~further facilitates the derivation of a tractable collapsed Gibbs--sampler to perform inference on the posterior distribution $p(\bar{\bPi},\bar{\bLambda}, \bz \mid \bY, \bc)$. 

Notice that, as mentioned before, when $\boldsymbol{c}$  is not available one can still implement the proposed ZIP--SBM~by simply replacing the  supervised Gnedin process prior $\textsc{gn}(\gamma; \bc)$ with the corresponding unsupervised version~$\textsc{gn}(\gamma)$. In this case, the resulting urn scheme representation coincides with the one in \eqref{eq8} after removing the factors $(n_{hc_v, -v}+\alpha_{c_v})/(n_{h,-v}+\alpha_0)$ and $\alpha_{c_v}/\alpha_0$.

\section{\Large 3. Bayesian Computation and Inference}\label{sec_3}
We derive here a collapsed Gibbs--sampler with a data--augmentation step to draw values from the intractable posterior distribution $p(\bar{\bPi},\bar{\bLambda}, \bz \mid \bY, \bc)$, and then leverage the simulated samples to perform Bayesian inference on $(\bar{\bPi},\bar{\bLambda}, \bz)$ under the ZIP--SBM model presented in \eqref{mod_tot1}--\eqref{mod_tot3}. To accomplish this goal, first notice that if the augmented data in the matrices $\bX$ and $\bW$ were known, then, recalling representation \eqref{eq1} of model \eqref{mod_tot1}, $p(\bar{\bPi},\bar{\bLambda}, \bz  \mid \bY,\bX,\bW,\bc)= p(\bar{\bPi},\bar{\bLambda}, \bz \mid \bX,\bW, \bc)=p(\bar{\bPi} \mid \bX,\bz)p(\bar{\bLambda} \mid \bW,\bz)p(\bz \mid \bX{,}\bW{,} \bc)$. Leveraging Beta--Bernoulli and Gamma--Poisson~conjugacy, together with representation  \eqref{eq1}, we have that
\begin{equation}
\begin{split}
(\bar{\pi}_{hk} \mid  \bX,\bz) &\sim \mbox{Beta}(a+x_{hk}, b+n_{hk}-x_{hk}), \qquad {\mbox{for every }} h=1, \ldots, H \ \mbox{and} \ k=1, \ldots, h, \\ \vspace{-5pt}
(\bar{\lambda}_{hk} \mid  \bW,\bz) &\sim \mbox{Gamma}(a_1+w_{hk}, a_2+n_{hk}), \quad \  \quad {\mbox{for every }} h=1, \ldots, H \ \mbox{and} \ k=1, \ldots, h,
\end{split}
 \label{eq9}
\end{equation}
where $n_{hk}$ is the total number of unique pairs involving a criminal in group $h$ and one in group $k$, while~$x_{hk}$ and $w_{hk}$ denote the sum, over such pairs, of the entries $x_{vu}$ and $w_{vu}$ in \smash{$\bX$ and $\bW$}, respectively. These~conjugacy properties also allow one to derive closed--form expressions for $p(\bX \mid \bz)$ and $p(\bW \mid \bz)$ that are useful to~compute $p(\bz \mid \bX,\bW, \bc)$. More specifically, exploiting conjugacy in \eqref{eq9} and recalling Section~\ref{sec_2.1}, we can marginalize out analytically $\bar{\bPi}$ and $\bar{\bLambda}$ in $p(\bX \mid \bz,\bar{\bPi})$ and $p(\bW \mid \bz,\bar{\bLambda})$, respectively, to obtain $p(\bX \mid \bz)$ and $p(\bW \mid \bz)$ as in \eqref{eq10}.

Combining \eqref{eq10} with the urn--scheme in \eqref{eq8}, it is therefore possible to derive closed--form expressions for the full--conditionals of the group assignment $z_v$ of each criminal $v=1, \ldots, V$ given those of the others $\bz_{-v}$, and the matrices $\bX$ and $\bW$. More specifically, by direct application of Bayes' rule, it follows that $(z_v \mid \bz_{-v}, \bX,\bW,\bc)$ is a categorical variable with full--conditional probabilities
\begin{equation}
 \mbox{pr}(z_v=h \mid \bz_{-v},\bX,\bW,\bc) \propto  \mbox{pr}(z_v=h \mid \bz_{-v},\bc)p(\bX \mid\bz_{-v}, z_v=h)p(\bW \mid\bz_{-v}, z_v=h),
 \label{eq11}
\end{equation}
for every $h=1, \ldots, H_{-v}+1$,  where $p(\bX \mid\bz_{-v}, z_v=h)$ and $p(\bW \mid\bz_{-v}, z_v=h)$ can be evaluated as in \eqref{eq10}, whereas $ \mbox{pr}(z_v=h \mid \bz_{-v},\bc)$ admits the closed--form expression from the urn scheme in  \eqref{eq8}.

As a consequence of the above derivations, it is possible to devise a simple Gibbs--sampler which iteratively simulates  the group assignment $z_v$ of every criminal $v$ from its tractable full--conditional~distribution. Iterating over $v=1, \ldots, V$, yields a Markov chain with stationary distribution~$p(\bz \mid \bX,\bW, \bc)$. Combining this result with \eqref{eq9} yields, therefore, a collapsed Gibbs--sampler~\citep{van2008partially} targeting $p(\bar{\bPi},\bar{\bLambda}, \bz \mid \bY,\bX,\bW, \bc)$.  

Despite its tractability, the above MCMC strategy requires $\bX$ and $\bW$ which are, in fact, not fully observed. Following \eqref{eq1} and Figure~\ref{figure:2}, the only available information is on $\bY=\bW \odot~({\bf 1}-\bX)$. Therefore, to implement the previously--derived scheme, it is necessary to introduce a data--augmentation step that generates $\bX$ and $\bW$. Leveraging  \eqref{eq5} and the representation \eqref{eq1} of model \eqref{mod_tot1}, an effective strategy to address such a goal would be~to sample from $p(\bX, \bW \mid \bY, \bar{\bPi},\bar{\bLambda}, \bz, \bc)=p(\bW \mid \bY, \bX,\bar{\bPi},\bar{\bLambda}, \bz)p(\bX \mid \bY,\bar{\bPi},\bar{\bLambda}, \bz)$, where $p(\bW \mid \bY, \bX,\bar{\bPi},\bar{\bLambda}, \bz)=\prod_{v=2}^V \prod_{u=1}^{v-1}p(w_{vu} \mid y_{vu}, x_{vu},\bar{\lambda}_{z_v,z_u})$  and $p(\bX \mid \bY,\bar{\bPi},\bar{\bLambda}, \bz)=\prod_{v=2}^V \prod_{u=1}^{v-1}p(x_{vu} \mid y_{vu},\bar{\pi}_{z_v,z_u},\bar{\lambda}_{z_v,z_u})$. As a direct consequence of the discussion in Section~\ref{sec_2.1}, samples from  $p(x_{vu} \mid y_{vu},\bar{\pi}_{z_v,z_u},\bar{\lambda}_{z_v,z_u})$ can be readily obtained by noticing~that
\begin{equation}
\begin{aligned}
\begin{cases} \scalemath{1}{(x_{vu} \mid y_{vu},\bar{\pi}_{z_v,z_u},\bar{\lambda}_{z_v,z_u}) \sim \delta_{0} } & \quad \text{if} \ \ y_{vu} > 0, 
 \\ \scalemath{1}{ (x_{vu} \mid y_{vu},\bar{\pi}_{z_v,z_u},\bar{\lambda}_{z_v,z_u}) \sim \mbox{Bern}[\mbox{pr}(x_{vu}=1 \mid y_{vu}=0,\bar{\pi}_{z_v,z_u},\bar{\lambda}_{z_v,z_u})]} & \quad \text{if} \ \ y_{vu} = 0,\\ \end{cases} 
\end{aligned}
 \label{eq12}
\end{equation}
independently for each $v=2, \ldots, V$ and $u=1, \ldots, v-1$, where $\delta_{0} $ is the Dirac delta at $0$, whereas $\mbox{pr}(x_{vu}=1 \mid y_{vu}=0,\bar{\pi}_{z_v,z_u},\smash{\bar{\lambda}_{z_v,z_u}})=\mbox{pr}(x_{vu}=1 \mid y_{vu}=0,z_v=h,z_{u}=k, \bar{\pi}_{hk}, \smash{\bar{\lambda}_{hk}})$ is the conditional probability derived in closed--form in \eqref{eq3}. Hence, if $y_{vu}>0$, no security strategy has been implemented and,~hence, $x_{vu}=0$. Conversely, if $y_{vu}=0$, such a zero may be either the result of a hiding mechanism or simply due to an actual zero tie that has not been obscured. Therefore, in this context $x_{vu}$ is drawn from the corresponding full--conditional Bernoulli variable. Given $x_{vu}$ and recalling again the discussion in Section~\ref{sec_2.1}, samples from  $p(w_{vu} \mid y_{vu}, x_{vu},,\bar{\lambda}_{z_v,z_u})$ can be generated as follows
\begin{equation}
\begin{aligned}
\begin{cases} \scalemath{1}{(w_{vu} \mid y_{vu}, x_{vu},\bar{\lambda}_{z_v,z_u}) \sim \delta_{y_{vu}} } & \quad \text{if} \ \ y_{vu} > 0, \\  
\scalemath{1}{(w_{vu} \mid y_{vu}, x_{vu},\bar{\lambda}_{z_v,z_u}) \sim \delta_{0} } & \quad \text{if} \ \ y_{vu} = 0 \ \mbox{and} \ x_{vu}=0,\\  
\scalemath{1}{(w_{vu} \mid y_{vu}, x_{vu},\bar{\lambda}_{z_v,z_u}) \sim \mbox{Poisson}(\bar{\lambda}_{z_v,z_u}) } & \quad \text{if} \ \ y_{vu} = 0 \ \mbox{and} \ x_{vu}=1,\\
\end{cases} 
\end{aligned}
 \label{eq13}
\end{equation}
independently for $v=2, \ldots, V$ and $u=1, \ldots, v-1$, where $\delta_{y_{vu}} $ is the Dirac delta at $y_{vu}$. Therefore, if $y_{vu}>0$ the tie has not been hidden and, therefore, the actual interaction $w_{vu}=y_{vu}$ has been~effectively observed.~Conversely, when $y_{vu}=0$ and $x_{vu}=0$, then $w_{vu}$ must be necessarily equal to zero, since, also in this case, no security~strategy has been implemented. Finally, if $x_{vu}=1$ then $y_{vu}=0$, and, as a consequence, $w_{vu}$ can be any value sampled from a $\mbox{Poisson}(\lambda_{vu}=\bar{\lambda}_{z_v,z_u})$. Combing \eqref{eq12}--\eqref{eq13} with the previously--derived scheme to sample from  $p(\bar{\bPi},\bar{\bLambda}, \bz \mid \bY,\bX,\bW,\bc)$ yields the data--augmentation collapsed Gibbs--sampler in Algorithm \ref{alg:gibbs}. Notice that such a routine samples, iteratively, from:
\begin{itemize}
\item{{\bf Step 1.} $p(\bX \mid \bY,\bar{\bPi},\bar{\bLambda}, \bz,\bc)=p(\bX \mid \bY,\bar{\bPi},\bar{\bLambda}, \bz)$,}
\vspace{2pt}
\item{{\bf Step 2. $p(\bW \mid \bY, \bX,\bar{\bPi},\bar{\bLambda}, \bz,\bc)=p(\bW \mid \bY, \bX,\bar{\bLambda}, \bz)$},}
\vspace{2pt}
\item{{\bf Step 3. $p(\bz \mid \bY, \bX,\bW,\bc)=p(\bz \mid \bX,\bW,\bc)$},}
\vspace{2pt}
\item{{\bf Step 4. $p(\bar{\bPi} \mid  \bY, \bX,\bW,\bar{\bLambda},\bz,\bc)=p(\bar{\bPi} \mid \bX,\bz)$},}
\vspace{2pt}
\item{{\bf Step 5. $p(\bar{\bLambda} \mid  \bY, \bX,\bW,\bar{\bPi},\bz,\bc)=p(\bar{\bLambda} \mid \bW,\bz)$}.}
\end{itemize}
Applying the results described in \citet{van2008partially}, it can be readily shown that Step 1.--Step 5. yield a Markov chain targeting the augmented posterior $p(\bar{\bPi},\bar{\bLambda}, \bz,\bX,\bW \mid \bY,\bc)$. In fact, Step 1.--Step 2. generate the augmented data from the joint full--conditional $p(\bX, \bW \mid \bY,\bar{\bPi},\bar{\bLambda}, \bz,\bc)$, whereas Step 3.--Step 5. sample from the parameters' full--conditional distribution  $p(\bar{\bPi},\bar{\bLambda}, \bz \mid \bY,\bX,\bW,\bc)$. 

Discarding the draws for $\bX$ and $\bW$ produced by Algorithm~\ref{alg:gibbs}, provides samples from the posterior $p(\bar{\bPi},\bar{\bLambda}, \bz \mid \bY, \bc)$ of interest. Leveraging these samples, we  then conduct posterior inference on redundancy structures encoded in $\bz$ via the variation of information (VI) framework introduced by \citet{wade2018} for Bayesian clustering.  The VI defines a proper  metric among partitions which computes distances between generic grouping structures through a comparison of  individual and joint entropies \citep[e.g.,][]{meilua2007comparing}. As discussed in \citet{wade2018}, such a metric yields a number of practical advantages over other popular alternatives, such as the Binder’s loss \citep{binder1978bayesian}, and   facilitates principled point estimation and uncertainty quantification directly within the space of partitions. In particular, under the VI approach, a point estimate for $\bz$ is obtained via $\hat{\bz}= \mbox{argmin}_{\bz'} \mathbbm{E}_{(\bz\mid \bY)}[\mbox{VI}(\bz,\bz')]$. Similarly, a $1-\alpha$ credible ball around $\hat{\bz}$ can be derived by collecting all the partitions with a VI distance from $\hat{\bz}$ less than a given threshold guaranteeing that the ball has at least  $1-\alpha$ posterior mass, while having minimum size possible. Such inference procedures are implemented via the \texttt{R} library \texttt{mcclust.ext}  after computing the $V \times V$ {\em posterior similarity (or co--clustering)} matrix $\bS$ whose generic element $s_{vu}$ estimates $\mbox{pr}(z_v=z_u \mid \bY)$ via the relative frequency of Gibbs samples in which \smash{$z_v^{(t)}=z_u^{(t)}$}. By relying solely on measures of posterior co--clustering, this approach to inference on $\bz$ does not suffer from possible label--switching issues and does not require relabeling strategies \citep{stephens2000dealing}. 

 \begin{algorithm}[t]
\caption{Data--augmentation collapsed Gibbs--sampler for ZIP--SBM}\label{alg:gibbs}
\begin{algorithmic}
\State --- {\bf input}: $\bY$ and prior hyperparameters $(a,b)$, $(a_1, a_2)$, and $\gamma$.
\State --- initialize $\bz$, $\bar{\bPi}$ and $\bar{\bLambda}$, and set the total number of MCMC iterations $\mbox{T}$.
\vspace{5pt}
\For{$t=1, \ldots, \mbox{T}$}
\For{$v=2 \ldots, V$ and $u=1, \ldots, v-1$}
    \State 1. sample $x_{vu}$  from \eqref{eq12}.
    \State 2. sample $w_{vu}$  from \eqref{eq13}.
       \EndFor
\For{$v=1 \ldots, V$}       
    \State 3.1 reorder the labels in $\bz_{-v}$  so that only groups $h=1, \ldots, H_{-v}$  are non--empty.
    \State 3.2 compute the full--conditional probabilities  for $z_v$ (up to a proportionality constant)  as in    \eqref{eq11}. The expressions  for the  
   \State  \hspace{4mm}  quantities   $p(\bX \mid\bz_{-v}, z_v=h)$ and $p(\bW \mid\bz_{-v}, z_v=h)$ in  \eqref{eq11} are  available  analytically as  in  \eqref{eq10}, whereas that for
      \State  \hspace{4mm}  $\mbox{pr}(z_v=h \mid \bz_{-v},\bc)$ can be   found in \eqref{eq8}.
         \State 3.3 sample $z_v$ from the categorical variable with probabilities obtained by normalizing  those  computed in 3.2. 
         \EndFor
         \State --- Let $H^{(t)}$ be the number of non--empty clusters in the previously--sampled $\bz$
         \For{$h=1 \ldots, H^{(t)}$ and $k=1, \ldots, h$}    
\State 4. sample $\bar{\pi}_{hk}$ from the full--conditional Beta distribution in \eqref{eq9}.
\State 5. sample $\bar{\lambda}_{hk}$ from the full--conditional Gamma distribution in \eqref{eq9}.
         \EndFor
\EndFor
\vspace{3pt}
\end{algorithmic}
\end{algorithm}    

Concerning inference on the block--specific interaction  rates and the zero--inflation probabilities, we emphasize that, although this objective would be possible by leveraging the simulated values of $\bar{\bPi}$ and $\bar{\bLambda}$ from Algorithm~\ref{alg:gibbs}, these quantities are associated with sampled partitions of $\bz$ which vary throughout the Gibbs--sampling routine. This makes inference less interpretable. In fact, in practice, it is more convenient to provide law enforcement~with a point estimate $\hat{\bz}$ of $\bz$, along with measures of uncertainty around such an estimate, and then study the plug--in posterior distribution $p(\bar{\bPi},\bar{\bLambda} \mid \bY,\hat{\bz})$ for $\bar{\bPi}$ and $\bar{\bLambda}$, given  $\hat{\bz}$. This objective can be readily accomplished by re--running Algorithm~\ref{alg:gibbs} without steps 3.1--3.3, and $\bz$ fixed at $\hat{\bz}$. This yields samples from $p(\bar{\bPi},\bar{\bLambda} \mid \bY,\hat{\bz})$ that can be used to obtain Monte Carlo estimates and uncertainty measures for the  block--specific interaction  rates and zero--inflation probabilities associated with the specific partition $\hat{\bz}$. Although this strategy does not propagate to $\bar{\bPi}$ and $\bar{\bLambda}$  the uncertainty within~$\bz$, as illustrated in Sections~\ref{sec_4} and \ref{sec_5}, the posterior for $\bz$ is often well--concentrated and, hence, the underestimation of uncertainty in $p(\bar{\bPi},\bar{\bLambda} \mid \bY,\hat{\bz})$, relative to $p(\bar{\bPi} ,\bar{\bLambda}\mid \bY)$, is typically negligible when compared with the gains in interpretability. Finally,~note that, due to the structured representation in \eqref{eq5}, estimation and uncertainty quantification on the tie--specific parameters matrices $({\bPi},{\bLambda})$ can be performed as a byproduct of the inference on $(\bar{\bPi},\bar{\bLambda}, \bz)$.

\vspace{5pt}

\section{\Large 4. Simulation Studies}\label{sec_4}
\begin{figure}[b]
\centering
    \includegraphics[trim=0cm 0cm 0cm 0cm,clip,width=1\textwidth]{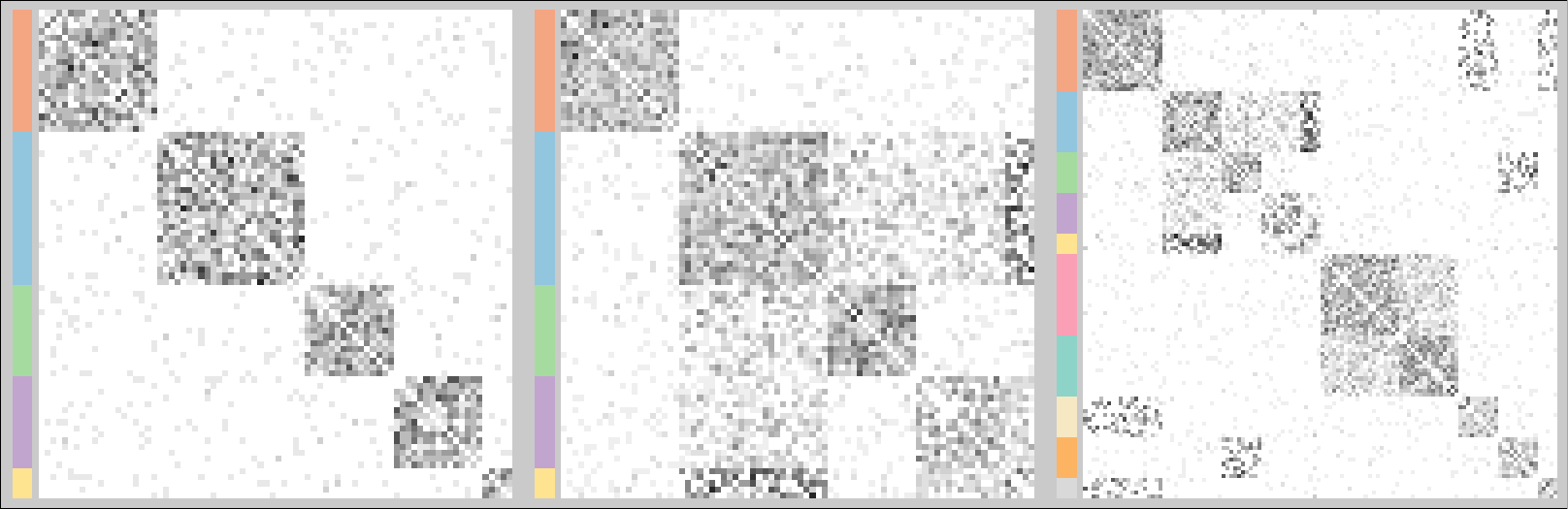}
                                \put(-425,164){{\bf Scenario 1}}
                                                \put(-260,164){{\bf Scenario 2}}
                \put(-100,164){{\bf Scenario 3}}
    \caption{\footnotesize{For scenarios 1, 2 and 3, graphical representation of the simulated adjacency matrix $\bY$. The color annotation of the rows displays the true grouping structure encoded in the true $\bz_0$ under each scenario. In the three adjacency matrices, the color of each entry  ranges from white to black as the corresponding tie goes from zero to the maximum observed count interaction. 
  }}
    \label{figure:3}
    \vspace{-10pt}
\end{figure}

To assess the performance of the proposed ZIP--SBM and illustrate its potential in inference on efficiency--security structures, we consider an in--depth analysis of three simulation scenarios based on different specifications of the parameters in model \eqref{mod_tot1}. As illustrated within Figure~\ref{figure:3},  the networks generated under these three scenarios are characterized by different sizes, number of clusters and various combinations of community, core--periphery and weakly--assortative structures, along with blocks displaying different sparsity and zero--inflation patterns. Some of these patterns reproduce those expected in criminal networks, thus providing a realistic assessment in the light of the motivating application. The three scenarios considered are described in detail below.

\begin{itemize}
\item {\bf Scenario 1:} Data are simulated from the generative representation \eqref{eq1} of the model in \eqref{mod_tot1} considering $V=80$  criminals partitioned into $H_0=5$ groups of size $n_1=20$, $n_2=25$, $n_3=n_4=15$ and $n_5=5$.  The matrix  $\bar{\bLambda}_0$ has diagonal entries equal to $3$ and off--diagonal ones set at $0.1$, whereas  $\bar{\bPi}_0$ encodes within--block zero--inflation probabilities of $0.05$ and across--block ones equal to $0.15$.  As illustrated within Figure~\ref{figure:3}, this yields a network $\bY$ characterized by five well--separated community structures and relatively mild security strategies. Notice that the setting $V=80$ is motivated by the attempt to assess the ZIP--SBM performance on a network with a size similar to the one considered in the 'Ndrangheta network application introduced within Section~\ref{sec_1.1}, where~the total number of criminals under analysis is equal to $84$.
\vspace{5pt}
\item {\bf Scenario 2:} We still  simulate data from representation \eqref{eq1} of model \eqref{mod_tot1}, focusing again on $V=80$  criminals allocated to \smash{$H_0=5$} different groups of size \smash{$n_1=20$, $n_2=25$, $n_3=n_4=15$ and $n_5=5$}.  However, in this case we consider more  heterogenous block--architectures beyond the simple communities, combined with a higher prevalence of covert structures. This goal is accomplished by considering larger values equal to $0.3$ for all the entries in $\bar{\bPi}_0$ except for $\bar{\pi}_{0; 2,1}=\bar{\pi}_{0; 1,2}$, $\bar{\pi}_{0; 3,1}=\bar{\pi}_{0; 1,3}$, $\bar{\pi}_{0; 4,1}=\bar{\pi}_{0; 1,4}$, $\bar{\pi}_{0; 5,1}=\bar{\pi}_{0; 1,5}$, $\bar{\pi}_{0; 4,3}=\bar{\pi}_{0; 3,4}$ and $\bar{\pi}_{0; 5,3}=\bar{\pi}_{0; 3,5}$ which are set to $0.6$, and $\bar{\pi}_{0; 1,1}$, $\bar{\pi}_{0; 2,2}$ and $\bar{\pi}_{0; 3,3}$ that we fix equal to $0$ for assessing the ability of ZIP--SBM in learning blocks with no sparsity due to security strategies. Similarly, the diagonal elements of $\bar{\bLambda}_0$ are set equal to $4$, while the off--diagonal ones coincide with $0.5$, except for $\bar{\lambda}_{0; 3,2}=\bar{\lambda}_{0; 2,3}$,  $\bar{\lambda}_{0; 4,2}=\bar{\lambda}_{0; 2,4}$ and $\bar{\lambda}_{0; 5,4}=\bar{\lambda}_{0; 4,5}$ that we set to~$2$, and $\bar{\lambda}_{0; 5,2}=\bar{\lambda}_{0; 2,5}$ which is equal to $6$.
\vspace{5pt}
\item {\bf Scenario 3:} The third scenario provides a more challenging setting which combines the first two --- still~under representation \eqref{eq1} of the model \eqref{mod_tot1} --- in a larger network among $V=120$ criminals  partitioned into $H_0=10$ groups of size $n_1=n_6=20$, $n_2=n_7=15$, $n_3=n_4=n_8=n_9=10$ and $n_5=n_{10}=5$. More~specifically, the $(5 \times 5)$--dimensional sub--matrices $\bar{\bPi}_{0[1:5,1:5]}$ and $\bar{\bLambda}_{0[1:5,1:5]}$ coincide with those specified in scenario 2, whereas  \smash{$\bar{\bPi}_{0[6:10,6:10]}$} and \smash{$\bar{\bLambda}_{0[6:10,6:10]}$} have the same entries as the zero--inflation probability matrix and rate matrix, respectively, in scenario 1, except for $\bar{\pi}_{0; 7,6}=\bar{\pi}_{0; 6,7}$ which is set to $0.05$, $\bar{\lambda}_{0; 7,6}=\bar{\lambda}_{0; 6,7}$ that is increased to $2$, and $\bar{\lambda}_{0; 6,6}$, $\bar{\lambda}_{0; 7,7}$ which we fix equal to $4$. In specifying \smash{$\bar{\bPi}_{0[1:5,6:10]}=\bar{\bPi}_{0[6:10,1:5]}^\intercal$} and \smash{$\bar{\bLambda}_{0[1:5,6:10]}=\bar{\bLambda}_{0[6:10,1:5]}^\intercal$} we consider instead zero--inflation probabilities and rates equal to $0.5$ and $0.2$, respectively, for all  the entries except
  $\bar{\pi}_{0; 10,1}=\bar{\pi}_{0; 1,10}, \bar{\pi}_{0; 8,1}=\bar{\pi}_{0; 1,8} $ and $\bar{\pi}_{0; 9,3}=\bar{\pi}_{0; 3,9}$ which are set to $0.6$, and $\bar{\lambda}_{0; 10,1}=\bar{\lambda}_{0; 1,10}, \bar{\lambda}_{0; 8,1}=\bar{\lambda}_{0; 1,8}$ and $\bar{\lambda}_{0; 9,3}=\bar{\lambda}_{0; 3,9}$ that we fix to $5$.
\end{itemize}

Figure~\ref{figure:3} provides a graphical illustration of the networks simulated under the three aforementioned scenarios. In analyzing the efficiency--security architectures underlying these three simulated networks we implement the newly--developed ZIP--SBM  in \eqref{mod_tot1}--\eqref{mod_tot3} along with two relevant competitors which  clarify the inference advantages associated with the proposed construction. These competitors include the recently--proposed supervised extended stochastic block model (ESBM) for binary networks \citep{legramanti2022extended} applied to a dichotomized version of the adjacency matrices in Figure~\ref{figure:3}, and an improved version of the Poisson SBM (P--SBM) in, e.g., \citet{mcdaid2013improved}, which leverages the supervised Gnedin process prior in Section~\ref{sec_2.2} on $\bz$, rather than the classical Dirichlet--multinomial. These choices guarantee a fair comparison among the three models, which rely on the same supervised prior on $\bz$, thus highlighting the benefits associated with the use of zero--inflated Poisson distributions for the observed ties, rather than Bernoulli or Poisson ones. 

Although inference on group structures can be accomplished also via alternative solutions, such as, for example, community detection algorithms \citep{girvan2002community,newman2006modularity,blondel2008fast} and spectral clustering  \citep{von2007tutorial}, the methodological and practical superiority of stochastic block models over these competing alternatives has been already illustrated in several studies; e.g., \citet{legramanti2022extended}. Due to this, we focus here on state--of--the--art SBM formulations that provide competitive alternatives to the proposed ZIP--SBM model in the context of criminal~networks. In particular, the ESBM with supervised Gnedin process prior represents a relevant competitor aligned with the routine practice of dichotomizing the weighted ties prior~to statistical modeling of criminal networks. As clarified below, this perspective yields substantial loss of information that could be avoided by  analyzing the observed ties on the original count scale. The Poisson SBM provides a sensible and popular solution which is, in fact, a degenerate case of the proposed ZIP--SBM~that~implicitly~assumes the lack of security structures. This assumption is not realistic in the context of criminal networks and, in fact, as illustrated in the following, the proposed ZIP--SBM not only expands the inference potentials of both ESBM and P--SBM, but also yields substantially more accurate reconstructions of the generative mechanisms underlying the three simulated networks.

Table~\ref{table_runtimes_MC} quantifies these advantages with a focus on posterior inference for the underlying grouping structures in the three simulation scenarios. Results for the ZIP--SBM, ESBM  and P--SBM are based on the same supervised Gnedin process prior for $\bz$ as in \eqref{mod_tot3}, with hyper--parameters $\gamma=0.3$ and  $\alpha_1= \ldots= \alpha_C=1$, where $C$~is~equal to $5$ in the first two scenarios, and to $10$ in the third. Supervision in this case is with respect to a contaminated version $\bc$ of the true group membership~labels in $\bz_0$. This is obtained by changing the true group allocation of 20 randomly--selected nodes, under each scenario, thereby allowing one to assess the ability to leverage informative exogenous partitions, while preserving robustness to possible noise and contamination in these external data. Albeit sharing the same prior on $\bz$, ZIP--SBM, ESBM  and P--SBM differ substantially in the likelihood for the~ties. More specifically, the ESBM relies on Bernoulli interactions with independent $\mbox{Beta}(1,1)$ --- i.e., uniform --- priors on the block probabilities \citep{legramanti2022extended}, whereas the P--SBM considers Poisson ties with  independent $\mbox{Gamma}(1,1)$ --- i.e., $\mbox{Exp}(1)$ --- priors for the block--specific interaction rates \citep[see e.g.,][]{mcdaid2013improved}. Such a prior is also considered for the entries of $\bar{\bLambda}$ in the proposed  ZIP--SBM. As for the hyper--parameters of the $\mbox{Beta}(a,b)$ priors on the zero--inflation probabilities in $\bar{\bPi}$, we rely instead on the more informative specification $(a=1,b=9)$ which is useful in facilitating a more conservative identification of security architectures. Although these structures are crucial for criminal organizations, it is realistic to expect that the efficiency--security tradeoff combined with advanced investigations progressively reduce the covert portions of the criminal network. These hyper--parameter settings always led to accurate inference on the partition structure and on the block--parameters, under several different network structures both in the simulation studies and in the application. Hence, we suggest $\gamma=0.3$, $\alpha_1= \ldots= \alpha_C=1$, $(a=1,b=9)$ and $(a_1=1,a_2=1)$ as a default choice.

Posterior inference under the ZIP--SBM relies on $10{,}000$ samples from Algorithm~\ref{alg:gibbs}, after a conservative burn--in of $10{,}000$.  A plain \texttt{R} implementation of such a routine in a standard laptop required 2.5 minutes to draw the total of $20{,}000$ samples in scenarios 1 and 2 ($V=80$, $H_0=5$). Under scenario 3  ($V=120$, $H_0=10$) such a runtime increased to $5.5$ minutes. Considering the running times of general Gibbs samplers in complex models such as this one, Algorithm~\ref{alg:gibbs} offers a rapid and effective implementation which can efficiently track most criminal networks. Convergence and mixing is monitored via the traceplots of the quantity $\mbox{VI}(\bz^{(t)},\bz_0)$, for $t=1, \ldots, \mbox{T}$, which, unlike other model--specific quantities, is available for both ZIP--SBM, ESBM and P--SBM, and informs not only on mixing and convergence of the MCMC chains for $\bz^{(t)}$, $t=1, \ldots, \mbox{T}$ under each of the three models, but also on the concentration of these chains around the true $\bz_0$. Although graphical analysis of these traceplots suggests much rapid convergence and effective mixing under ZIP--SBM in all the three scenarios, we opt for a conservative burn--in which can be safely employed also for the competing ESBM and P--SBM. In fact,  in the second and third scenario, the P--SBM experiences  challenges in convergence, thus requiring the conservative burn--in employed. Posterior inference under ESBM proceeds via the Gibbs--sampler in \citet{legramanti2022extended}, while  P--SBM can be implemented via a minor modification of  Algorithm~\ref{alg:gibbs} removing Steps 1, 2 and 4, and setting $\bY=\bW$. As in \citet{legramanti2022extended}, we also found results robust to moderate changes~in the  Gnedin process hyper--parameter  $\gamma \in (0,1)$. In particular, setting $\gamma$ to either $0.1$ or $0.7$, rather than $0.3$, did not change the ZIP--SBM performance displayed in Table~\ref{table_runtimes_MC}. The same robustness to the choice of the prior hyper--parameters  was observed also when considering $(a=1,b=6)$ or $(a=1,b=4)$, instead of $(a=1,b=9)$. Similarly, setting $(a_1,a_2)$ to either $(a_1=1,a_2=2.5)$  or $(a_1=2.5,a_2=1)$, rather than $(a_1=1,a_2=1)$,  did not modify the results in Table~\ref{table_runtimes_MC} for ZIP--SBM, except for the third scenario where the hyper--parameter setting $(a_1=2.5,a_2=1)$~led the ZIP--SBM to collapse three groups, out of the ten in total, into a single one. This slight deterioration in clustering accuracy is, however, not substantial, provided that the three~collapsed~groups~are characterized by highly--similar  block--parameters, and thus, inference on the underlying data--generating mechanism is not substantially affected. All routines proved robust also to the initialization of the partition structure $\bz$. Although initializing the sampling algorithms to $V$ singleton groups --- with each node occupying its own cluster --- generally led to a slightly--more--rapid convergence, other starting configurations based on randomly--generated node partitions or more extreme settings with all the nodes allocated to the same unique group, did not affect  inference.

\setlength{\tabcolsep}{6.5pt}
\begin{table*}[t]
\renewcommand{\arraystretch}{1}
\centering
\caption{\footnotesize{For scenarios 1, 2 and 3, performance of ZIP--SBM and relevant competitors (ESBM and  P--SBM) in recovering the true underlying partition $\bz_0$.  This performance  is measured via: (i) the number \smash{$\hat{H}$} of distinct clusters in the estimated grouping structure $\bf{\hat{z}}$, (ii)  the VI distance $\mbox{VI}(\hat{\bz},\bz_0)$ between the estimated partition $\hat{\bz}$ and the true one $\bz_0$, (iii) the normalized mutual information $\mbox{NMI}(\hat{\bz},\bz_0)$ between  $\hat{\bz}$ and  $\bz_0$, (iv) the posterior mean $\mathbbm{E}[\mbox{VI}(\bz,\bz_0) \mid \bY]$ of the VI distance from the true  $\bz_0$ (which measures the overall concentration of the posterior for $\bz$ around the true underlying partition $\bz_0$), and (v) the distance $\mbox{VI}(\hat{\bz},\bz_b)$ between the estimated partition $\hat{\bz}$ and the $95\%$ credible bound $\bz_b$. Bold values denote best performance in each column.}}
\vspace{5pt}
\label{table_runtimes_MC}
\begin{adjustbox}{width=1\textwidth,center=\textwidth}
\begin{tabular}[c]{l|ccc|ccc|ccc|ccc|ccc}
 \multicolumn{1}{c}{}&   \multicolumn{3}{c}{$\hat{H}$}  & \multicolumn{3}{c}{$\mbox{VI}(\hat{\bz},\bz_0)$} & \multicolumn{3}{c}{ $\mbox{NMI}(\hat{\bz},\bz_0)$} &  \multicolumn{3}{c}{$\mathbbm{E}[\mbox{VI}(\bz,\bz_0) \mid \bY]$}  & \multicolumn{3}{c}{$\mbox{VI}(\hat{\bz},\bz_b)$}   \\ 
\midrule
  \textsc{Scenario} & 1 & 2 &3 & 1 & 2&3 & 1 & 2&3&1&2&3&1& 2&3 \tabularnewline[0.5em]
   \hline 
ZIP--SBM& {\bf 5} &    {\bf 5}&{\bf 10}&  {\bf 0.00} &   {\bf 0.00} &{\bf 0.00}&  {\bf 1.00} &   {\bf 1.00} &{\bf  1.00}&  {\bf 0.0009} &  {\bf 0.0001}& ${\bf 0.0001}$&  {\bf 0.00}&   {\bf 0.00}& {\bf 0.00} \\
\hline
ESBM& {\bf 5} &   4&8&  {\bf 0.00} & 0.20 &0.40&  {\bf 1.00} &  0.91 &0.87&  0.0027& 0.2156& $ 0.4036$&  {\bf 0.00}&  0.13& {\bf 0.00}\\
P--SBM&  {\bf 5} &   8&13&   {\bf 0.00} &  0.45 &0.23&   {\bf 1.00} &  0.84 &0.93&  0.0024 & 0.5880& $ 0.2491$&   {\bf 0.00}&  0.69& 0.09\\
\midrule
\end{tabular}
\end{adjustbox}
\end{table*}

\setlength{\tabcolsep}{3.5pt}
\begin{table*}[b]
\renewcommand{\arraystretch}{1}
\centering
\caption{\footnotesize{Performance of the ZIP--SBM in 100 replicated studies from scenarios 1, 2 and 3.  This performance is measured via the median of the quantities in  Table \ref{table_runtimes_MC}, computed over the 100 replicated simulations from each of the three scenarios. The values within brackets correspond to the difference between the $90\%$ and $10\%$ quantiles of the different measures over the 100 replicates. These latter values quantify the variability of the performance measures under analysis across the 100 simulations.}}
\vspace{5pt}
\label{table_replicated}
\begin{adjustbox}{width=1\textwidth,center=\textwidth}
\begin{tabular}[c]{l|ccc|ccc|ccc|ccc|ccc}
 \multicolumn{1}{c}{}&   \multicolumn{3}{c}{$\hat{H}$}  & \multicolumn{3}{c}{$\mbox{VI}(\hat{\bz},\bz_0)$} & \multicolumn{3}{c}{ $\mbox{NMI}(\hat{\bz},\bz_0)$} &  \multicolumn{3}{c}{$\mathbbm{E}[\mbox{VI}(\bz,\bz_0) \mid \bY]$}  & \multicolumn{3}{c}{$\mbox{VI}(\hat{\bz},\bz_b)$}   \\ 
\midrule
  \textsc{Scenario} & 1 & 2 &3 & 1 & 2&3 & 1 & 2&3&1&2&3&1& 2&3 \tabularnewline[0.5em]
   \hline 
ZIP--SBM& 5.0 &   5.0& 10.0&  0.00 &   0.00 & 0.07&  1.00 & 1.00 &0.99& 0.0005  &  0.0002& 0.0705&   0.00&   0.00&  0.00\\
& (0.00) &   (0.00)& (2.00)&  (0.00) &   (0.00) & (0.24)&  (0.00) & (0.00) &(0.07)&  (0.002) &  (0.003)& (0.275)&  (0.00)&   (0.00)& (0.20)\\
\midrule
\end{tabular}
\end{adjustbox}
\end{table*}

Leveraging the posterior samples for $\bz$ from the implementations of the ZIP--SBM, ESBM and P--SBM under the three simulation scenarios, Bayesian inference on the partition structure proceeds under the VI framework discussed in detail in Section~\ref{sec_3}. Results of these  analyses are illustrated in Table~\ref{table_runtimes_MC}, through several performance measures which quantify the accuracy of the estimated partition $\hat{\bz}$ in recovering the true one ${\bz}_0$ (see $\hat{H}$, $\mbox{VI}(\hat{\bz},\bz_0)$ and $\mbox{NMI}(\hat{\bz},\bz_0)$), along with the concentration of the entire posterior for $\bz$ around $\bz_0$ (see $\mathbbm{E}[\mbox{VI}(\bz,\bz_0) \mid \bY]$), and the size of the credible ball associated with such a posterior (see $\mbox{VI}(\hat{\bz},\bz_b)$). Notice that, for completeness, we also compute the normalized mutual information $\mbox{NMI}(\hat{\bz},\bz_0) \in [0,1]$ between  $\hat{\bz}$ and  $\bz_0$ which provides a measure of similarity among the two partitions \citep[e.g.,][]{newman2020}. Large values of $\mbox{NMI}(\hat{\bz},\bz_0)$~imply~low~$\mbox{VI}(\hat{\bz},\bz_0)$.~As outlined in Table~\ref{table_runtimes_MC}, all these measures confirm the superior performance of the ZIP--SBM in accurately learning the true underlying partition $\bz_0$ in all scenarios. Despite yielding a slightly lower concentration of the posterior around $\bz_0$ than the ZIP--SBM, both the ESBM and~P--SBM achieve a similarly--accurate performance in point estimation within scenario 1. Conversely, in scenarios 2 and 3 the inference accuracy of both the ESBM and~P--SBM deteriorates, while the ZIP--SBM remains still able to recover exactly the true underlying partition  in $\bz_0$. The failure of the ESBM is caused by the information loss arising from dichotomization, which makes the fourth and fifth groups indistinguishable in scenario 2.  The same holds for clusters four--five and six--seven in scenario 3. The challenges encountered by the P--SBM are instead attributable to the inability of accounting for the sparsity in scenarios 2 and 3, which~results in the creation of additional groups to cope with the lack of flexibility in the  Poisson likelihood. The ZIP--SBM  avoids data--dichotomization and properly incorporates both sparsity and weighted ties information, thereby achieving exact recovery of $\bz_0$ and effective concentration of the posterior distribution around such a $\bz_0$  also in scenarios~2 and 3.

As displayed in Table~\ref{table_replicated}, the high accuracy of the  ZIP--SBM in Table~\ref{table_runtimes_MC} is not specific to the three simulated networks in Figure~\ref{figure:3}, but rather holds generally in replicated simulations from the generative mechanisms behind the three scenarios. More specifically, we replicate the analyses in Table~\ref{table_runtimes_MC} --- with focus on the proposed ZIP--SBM --- for 100 different networks simulated under each of the three scenarios, thereby obtaining, for every~scenario, 100 different values of the performance measures reported in Table~\ref{table_runtimes_MC} for the ZIP--SBM.  Table~\ref{table_replicated} displays~the~median of these 100 values along with the difference between the $90\%$ and $10\%$ quantiles to quantify a range of variation across the 100 simulations. These results further strengthen the proposed ZIP--SBM which proves highly reliable and yields accurate inference on the true grouping structures among the nodes in almost all the 100 replications, under each  of the three simulation scenarios. 

Recalling the results in Table~\ref{table_runtimes_MC}, notice that, even if the ESBM and P--SBM achieve accurate point estimation of $\bz$ in some~scenario (e.g., scenario 1), these two models are, by construction, not able to disentangle security--efficiency architectures, and learn accurately both $\bar{\bPi}$ and $\bar{\bLambda}$. This  is clear when comparing the~posterior mean of $\bar{\bPi}$ and $\bar{\bLambda}$ associated with the estimated partition $\hat{\bz}$ under the ZIP--SBM in scenario 1, with those provided~by the ESBM and  P--SBM, respectively. More specifically, a proxy for $\bar{\bPi}$ under the ESBM~can~be~obtained~as~the posterior mean of the block--specific probabilities of a zero tie, while $\bar{\bLambda}$ can be compared against the posterior mean of the Poisson rates estimated, for each block, via the P--SBM. Under the ZIP--SBM, the mean absolute difference between the entries in $\bar{\bPi}_0$ and those in the estimated $\bar{\bPi}$ is $0.03$, a value which is orders of magnitude lower than the overall error of $0.52$ achieved under the ESBM. Similarly, the absolute error in recovering~$\bar{\bLambda}_0$ under the ZIP--SBM is $0.03$, again improving over the P--SBM error, which is $0.05$.~Hence,~although~the~ESBM and P--SBM recover $\bz$ in scenario~1, these models are not as effective as the ZIP--SBM in estimating~$\bar{\bPi}_0$~and~$\bar{\bLambda}_0$. The poor performance of the ESBM~in learning $\bar{\bPi}_0$ is due to the fact that, by construction, such a model cannot separate truly--zero ties from hidden~ones. Similarly,~the P--SBM assumes that all the zero--ties are realizations~from a Poisson, and hence, yields an underestimation of the actual rates of interaction in the different blocks. The ZIP--SBM not only can learn both $\bar{\bPi}$ and $\bar{\bLambda}$, but also provides relatively accurate estimates of both matrices. These are further confirmed by the posterior standard deviations for the entries in $\bar{\bPi}$ and $\bar{\bLambda}$ whose first, second and third quartiles are $[0.088, 0.102, 0.108]$ and $[0.024, 0.030, 0.041]$, respectively. This high accuracy in inference for the entries of $\bar{\bPi}$ and $\bar{\bLambda}$ is preserved also for blocks displaying no sparsity patterns induced by zero inflations. For example, in scenario 2, we have $\bar{\pi}_{0; 1,1}=\bar{\pi}_{0; 2,2}=\bar{\pi}_{0; 3,3}=0$ and $\bar{\lambda}_{0; 1,1}=\bar{\lambda}_{0; 2,2}=\bar{\lambda}_{0; 3,3}=4$. The estimates provided by the ZIP--SBM for these parameters are \smash{$\hat{\bar{\pi}}_{1,1}=0.007$, $\hat{\bar{\pi}}_{2,2}=0.006$, $\hat{\bar{\pi}}_{3,3}=0.012$ and $\hat{\bar{\lambda}}_{1,1}=3.9$, $\hat{\bar{\lambda}}_{2,2}=3.9$, $\hat{\bar{\lambda}}_{3,3}=4.4$}.

\begin{figure}[b]
\centering
    \includegraphics[trim=0cm 0cm 0cm 0cm,clip,width=1\textwidth]{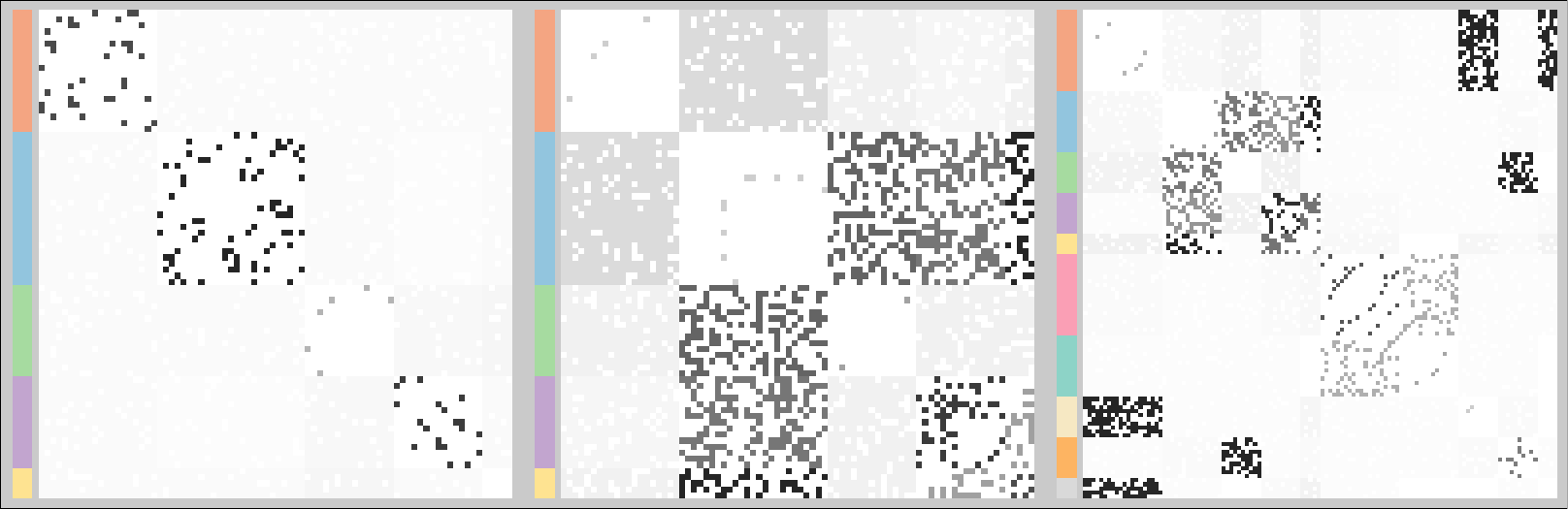}
                                \put(-425,164){{\bf Scenario 1}}
                                                \put(-260,164){{\bf Scenario 2}}
                \put(-100,164){{\bf Scenario 3}}
    \caption{\footnotesize{For scenarios 1, 2 and 3, graphical representation of the estimated probabilities that each observed zero tie $y_{vu}=0$ corresponds, in fact, to a strictly positive obscured interaction. The color annotation of the rows displays the grouping structure $\hat{\bz}$ estimated under the proposed ZIP--SBM model. The color of each entry  in the three matrices ranges from light gray to black as the estimated probabilities, under the   proposed ZIP--SBM, range from 0 to 1. White entries denote non--zero observed ties ($y_{vu}>0$).
}}
    \label{figure:4}
\end{figure}

As clarified in Figure~\ref{figure:4}, the above advantages open avenues for a new set of inference strategies that~cannot be considered under the ESBM and  P--SBM, but are of key interest in criminology. A core one is~the~quantification of which observed zero ties $y_{vu}=0$ are, in fact, due to a security strategy applied to a non--zero underlying interaction among the generic criminals $v$ and $u$. Following Section \ref{sec_2.1}, these events have probabilities~defined in~\eqref{eq31} for each $v=2, \ldots, V$ and $u=1, \ldots, v-1$, which can be evaluated at the true parameters $(\bz_0,\bar{\bPi}_0,\bar{\bLambda}_0)$,~and also estimated via Monte Carlo by averaging~\eqref{eq31} over the posterior samples  from $p(\bar{\bPi},\bar{\bLambda} \mid \bY, \hat{\bz})$ produced by the Algorithm~\ref{alg:gibbs}, in combination with~$\hat{\bz}$. Under the ZIP--SBM, the absolute difference between these~true and estimated probabilities, averaged over the different pairs of nodes, is $0.02$, $0.15$ and $0.10$ for scenarios 1, 2 and 3, respectively. This result supports the overall accuracy of the ZIP--SBM in the estimation of such probabilities, thereby allowing one to unveil which zero ties should be prioritized~in~the~investigations.~Figure~\ref{figure:4}~illustrates~the output of these analyses under the ZIP--SBM for the three networks in Figure~\ref{figure:3}. According to~Figure~\ref{figure:4},~among the observed zero ties in the three networks in Figure~\ref{figure:3}, the estimates provided by the ZIP--SBM correctly prioritize those belonging to blocks in which the associated rate and zero--inflation probability make such zero ties highly unusual under a standard Poisson.  This effectiveness is also confirmed by the posterior standard~deviations for the probabilities under analysis, whose first, second and third quartiles are $[0.015,  0.017, 0.021]$,  $[0.045, 0.057, 0.075]$, and $[0.014, 0.023, 0.053]$ for scenarios 1, 2 and 3, respectively, suggesting accurate concentration under the~ZIP--SBM. It is important to emphasize that, in the more sparse blocks within scenarios 2--3, the estimation~accuracy slightly deteriorates due to the lack of enough non--zero ties to effectively infer the rate of the count distribution. This is a general problem in zero--inflated contexts. Nonetheless, the results of the ZIP--SBM are still satisfactory even~for~these challenging scenarios where state--of--the--art alternatives  fail.

\section{\Large 5. Evidence of Efficiency and Security Structures in the  Infinito Network}\label{sec_5}
We conclude by illustrating the innovative inference potential of the proposed ZIP--SBM in the analysis of~the {\em Infinito} 'Ndrangheta network presented  in Section~\ref{sec_1.1}. State--of--the--art analyses of such a network focus either on detecting basic community architectures that suggest an overly--simplified horizontal structure underlying~the 'Ndrangheta organization \citep[e.g.,][]{calderoni2017communities}, or study a dichotomized version of the original weighted network  \citep{legramanti2022extended}  which hinders the potential to unveil those security and efficiency architectures of key interest in criminology \citep[][]{morselli2007efficiency,calderoni2012structure,Bouchard,cavallaro2020disrupting}. In fact, as illustrated in the simulation studies in Section~\ref{sec_4}, neither the ESBM, nor P--SBM, can disentangle these two fundamental structures. Conversely, the proposed ZIP--SBM is inherently motivated by the attempt~to address such an endeavor via a principled model that can effectively incorporate and quantify both security and efficiency strategies, thus motivating our focus on the ZIP--SBM in the {\em Infinito} network study.

\begin{figure}[b]
\centering
    \includegraphics[trim=0cm 0cm 0cm 0cm,clip,width=0.9\textwidth]{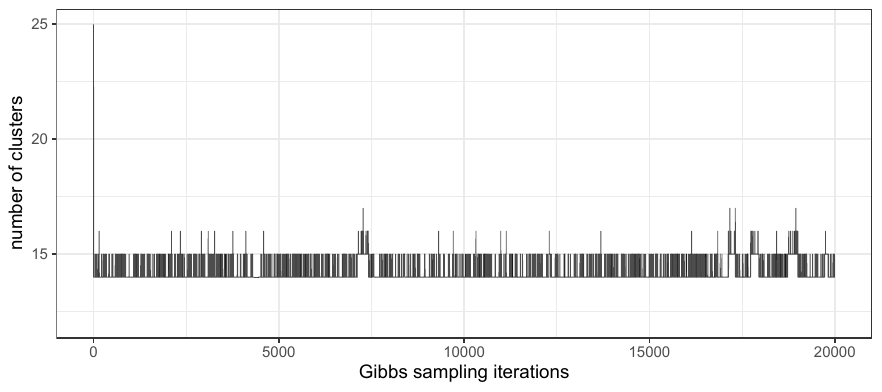}
      \vspace{-10pt}
    \caption{\footnotesize{For the {\em Infinito} 'Ndrangheta network application, traceplot of the number~$H^{(t)}$~of~occupied~clusters~in~the~sampled~partition~$\bz^{(t)}$, for $t=1, \ldots, 20{,}000$, under Algorithm~\ref{alg:gibbs}.}
  \vspace{-5pt}}
    \label{figure:traceplot}
\end{figure}

\begin{figure}[t]
\centering
    \includegraphics[trim=2.1cm 2.5cm 1.1cm 2.3cm,clip,width=0.85\textwidth]{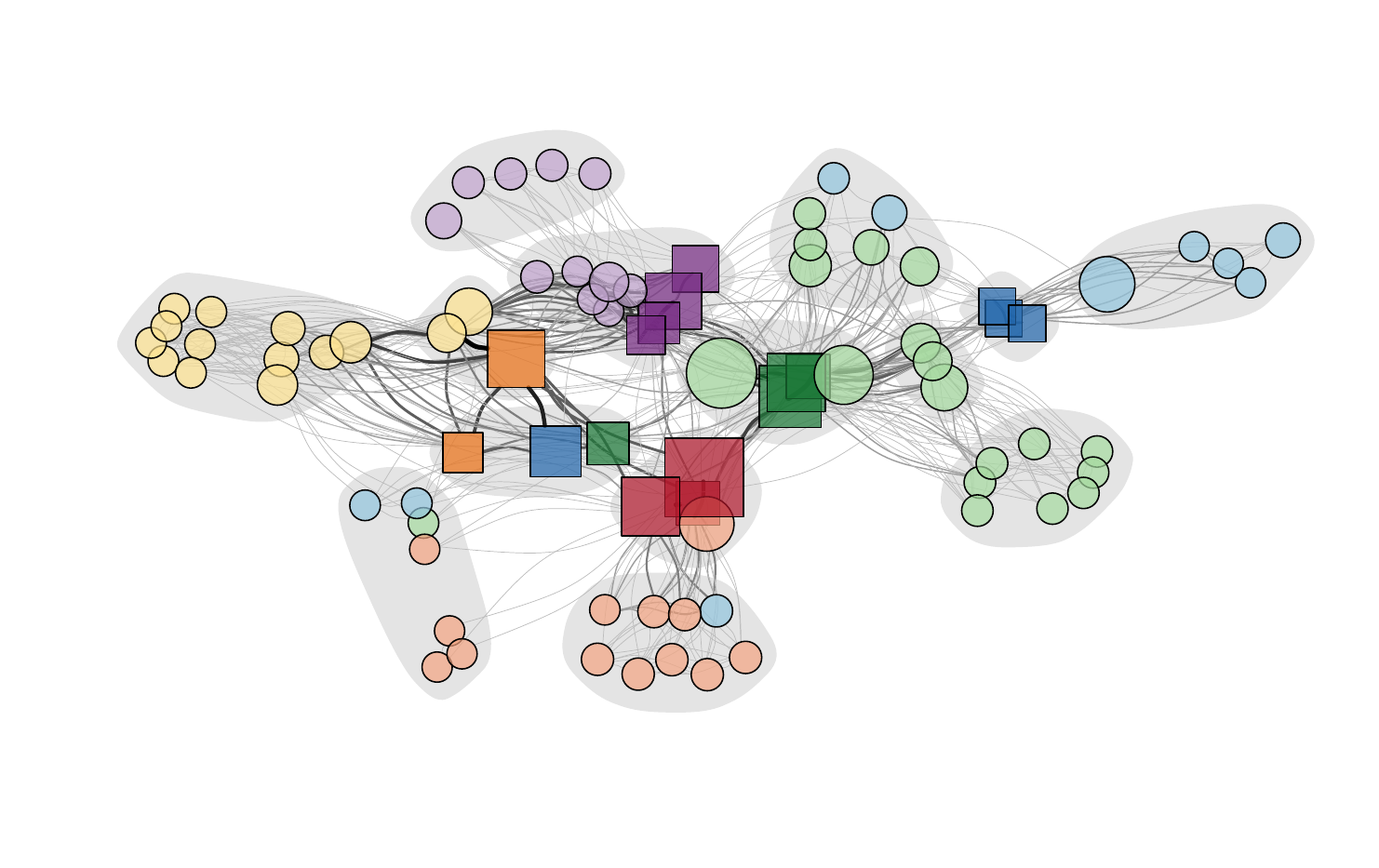}
      \vspace{-10pt}
    \caption{\footnotesize{Graphical representation of the {\em Infinito}  network along with the grouping structure estimated under the proposed ZIP--SBM. The positions of the different criminals are obtained via force directed placement \citep{fru1991}, whereas colors indicate the presumed {\em locale} membership and role (darker square nodes indicate the suspected bosses of each {\em locale}, while lighter circles denote simple affiliates). The size of each node is proportional to its betweenness, while the gray areas highlight the different groups encoded in $\hat{\bz}$.}
  \vspace{-5pt}}
    \label{figure:5}
\end{figure}

To support the above remark, we analyze the  {\em Infinito} network under the  proposed ZIP--SBM,~considering~the same hyper--parameters and MCMC settings as in the simulation studies in Section~\ref{sec_4}. This is useful in order~to check the robustness of results when such default choices are considered in the analysis of substantially different networks. In fact, we still obtain satisfactory convergence and mixing  (see e.g., Figure~\ref{figure:traceplot}), along with a highly--informative reconstruction of the redundancy structures underlying~the  {\em Infinito} network. These group patterns, encoded in $\hat{\bz}$, are shown in Figure~\ref{figure:5} which not only yields quantitative support to a number~of criminology~theories, but also unveils more nuanced dynamics underlying this specific criminal organization. The former are evident in a preference to create redundancies  within {\em locali}, rather than across these modular units, while further diversifying the affiliates from bosses in the formation of groups. This suggests a peculiar ability~to guarantee resilience at different levels of the organization, while clarifying that the creation of redundancies is more challenging when moving from peripheral groups of simple affiliates to the core ones, mostly comprising bosses with highly specific human and social capital. In fact, such central groups have progressively lower~size.

Although the above recurring patterns suggest a highly--regulated organizational structure, which is known to characterize 'Ndrangheta \citep[e.g.,][]{paoli2007mafia,catino2014mafias,sergi2016ndrangheta}, the flexibility of the proposed   ZIP--SBM is also able to detect more peculiar dynamics related to the history of the specific organization reported in the judicial documents of ``Operazione {\em Infinito}''. In fact, although the supervised Gnedin process prior in \eqref{mod_tot3} facilitates some overlap between the exogenous partition $\bc$ provided by role--{\em locale} information and the endogenous one $\hat{\bz}$ induced by the redundancies within the network, Figure~\ref{figure:5} still highlights peculiar interaction dynamics among criminals in different \emph{locali}. For example, according to the judicial documents, the criminal~belonging~to~the blue {\em locale} that has been allocated to the group of simple affiliates from the red one, is a member trying to create a new {\em locale}. To this end, the results in Figure~\ref{figure:5} suggest that such a criminal might be in the process of recruiting affiliates from outside the area of its original {\em locale}, with a preference for those belonging to the red one. It is also interesting to notice a group composed of bosses from different {\em locali}, with two members relatively far, under a network perspective, from the  corresponding original {\em locali}. This is in line with the fact that these two bosses supported an unsuccessful attempt to increase the independence of the ‘Ndrangheta group in Lombardy~from~the leading Calabria families, causing a need for these members to move away from their original areas and strengthen relations with other~{\em locali}. From this perspective, Figure~\ref{figure:5} points toward a progressive movement in the direction of the yellow {\em locale}. It is also worth noticing that this independence attempt led to the murder of a high--rank member that undermined the stability of the green locale {\em locale}. Figure~\ref{figure:5}  confirms that this event has resulted in a fragmentation of the green {\em locale} in multiple sub--groups.~Finally, notice that the core~groups~of~bosses~also comprise some simple affiliates, thus suggesting that the corresponding role might be more central than reported~in the judicial documents.  This is the case of the central affiliate allocated to the group of bosses from the green~{\em locale}, who is, in fact, a high--rank member of the organization~with~key~coordinating~roles~among~\emph{locali}.

\begin{figure}[t]
\centering
    \includegraphics[trim=0cm 0cm 0cm 0cm,clip,width=0.9\textwidth]{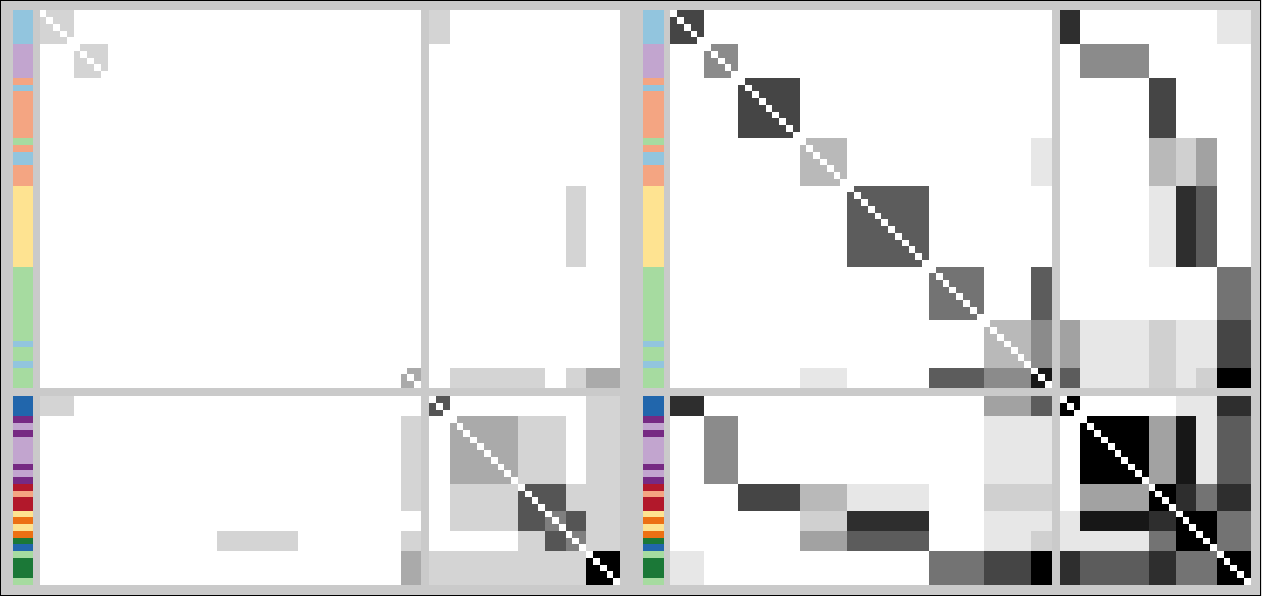}
                \put(-155,214){{\bf Efficiency structure}}
                                \put(-370,214){{\bf Security structure}}
    \caption{\footnotesize{Adjacency matrices representing security and efficiency block structures of the {\em Infinito} network, inferred under the proposed ZIP--SBM. Criminals are re--ordered
and partitioned in blocks according to the estimated grouping structure $\hat{\bz}$. Side colors correspond to the different {\em locali}, with darker and lighter shades denoting bosses and affiliates, respectively. The security structure displays, for each block, the posterior mean of the probability that a generic observed zero tie in that block is the result of a hiding mechanism. Conversely, the efficiency structure represents, for each block, the posterior mean of the probability that a generic tie within that block, either hidden or not, is $>0$. The color of each entry in the two matrices ranges from white to black as the corresponding probabilities go from $0$ to $1$. The gray lines separate groups containing only simple affiliates from those comprising also bosses.
  \vspace{0pt}}}
    \label{figure:6}
\end{figure}
\begin{figure}[t!]
\vspace{-1pt}
\centering
    \includegraphics[trim=0cm 0cm 0cm 0cm,clip,width=0.9\textwidth]{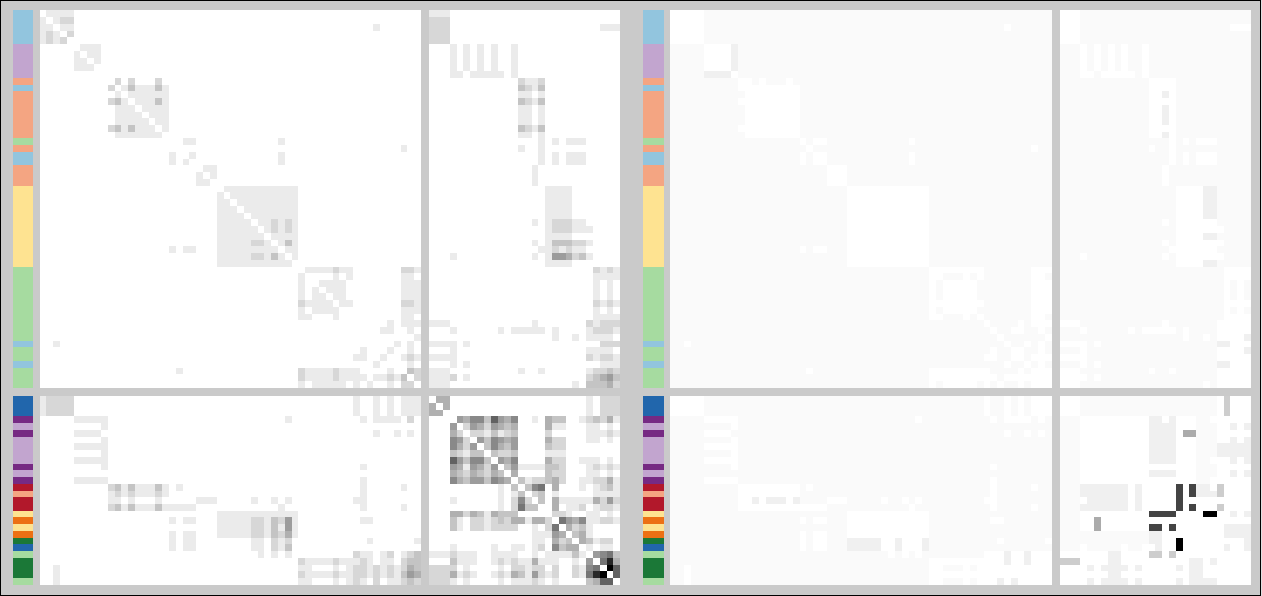}
         \put(-165,214){{\bf Zero ties to be investigated}}
                                \put(-370,214){{\bf Observed network}}
    \caption{\footnotesize{Graphical representation of the adjacency matrix $\bY$ associated with the {\em Infinito} network, along with the posterior means of the probabilities that each observed zero tie $y_{vu}=0$ corresponds, in fact, to a strictly positive obscured interaction.  Criminals are re--ordered and partitioned in blocks according to the  group structure estimated under ZIP--SBM. Side colors correspond to the {\em locali}, with darker and lighter shades denoting bosses and affiliates, respectively. In the first matrix, the color of each entry  ranges from white to black as the corresponding tie goes from zero to the maximum observed count interaction, whereas in the second  the color goes from light gray to black as the estimated probabilities, under the  ZIP--SBM, range from 0 to 1. White entries denote non--zero observed ties ($y_{vu}>0$). Gray lines separate groups containing only simple affiliates from those comprising also bosses.
  \vspace{-9pt}}}
    \label{figure:7}
\end{figure}

The above results, which are based on the minimum--VI point estimate  $\hat{\bz}$ of $\bz$, are further supported by the fact that the whole posterior distribution is well--concentrated around $\hat{\bz}$. More specifically, the~VI distance between $\hat{\bz}$ and the partition $\bz_b$ at the edge of the $95\%$ credible ball is $0.301$, a value much lower than the maximum achievable VI distance between two generic partitions of the $V=84$ analyzed criminals, which is  $\log_2 84 =6.392$  \citep{wade2018}. Similarly, the first and third quartiles of the posterior distribution on the number of non--empty groups are both equal to $14$, which coincides with the total number $\hat{H}=14$  of occupied groups in $\hat{\bz}$. These results motivate an in--depth analysis of the covert and overt architectures behind the {\em Infinito} network leveraging the strategies presented in Sections~\ref{sec_2.1} and \ref{sec_3}, which condition on the estimated $\hat{\bz}$. 

Figure~\ref{figure:6} illustrates the innovative inference potentials of the proposed ZIP--SBM on these key, yet--unexplored, architectures. In fact, while the analysis of  \citet{legramanti2022extended} on the dichotomized version of the  {\em Infinito} network yields a similarly--refined reconstruction of the hierarchical group structures underlying the targeted 'Ndrangheta organization, the perspective considered by the authors rules out the possibility of disentangling and quantifying security and efficiency structures. Figure~\ref{figure:6} addresses this objective via a graphical representation of the posterior mean for the probabilities in \eqref{eq3} and \eqref{eq4}, respectively, which can be computed, for each $v=2, \ldots, V$ and $u=1, \ldots, v-1$, through Monte Carlo leveraging $\hat{\bz}$ and the posterior samples from $p(\bar{\bPi},\bar{\bLambda} \mid \bY, \hat{\bz})$; see also Section~\ref{sec_3} for further details and recall the definition of $\pi_{vu}$ and $\lambda_{vu}$ in \eqref{eq5}. As discussed in Section~\ref{sec_2.1}, the first probability quantifies the chance that a potential zero tie  among criminals $v$ and $u$ is the result of a security strategy. Conversely, the second evaluates the probability that $v$ and $u$ establish a non--zero tie,~irrespectively of whether this tie has been obscured or not, thus unveiling the actual efficiency structures of the organization.

As shown in Figure~\ref{figure:6}, the ZIP--SBM uncovers a hierarchical efficiency structure underlying~the {\em Infinito} network, characterized by a peculiar combination of community architectures with across--group~interactions that occur through highly--structured core--periphery patterns between the groups of affiliates and those including the bosses. The within--block interactions progressively intensify while~moving from the peripheral groups of affiliates to~the core ones of bosses, whereas the across--block~patterns clearly suggest a tendency of simple affiliates to connect to the core only via the bosses of the corresponding {\em locale}. These findings provide important quantitative support~to qualitative criminology theories on the hierarchical structure of 'Ndrangheta  \citep[see e.g.,][]{paoli2007mafia,catino2014mafias,sergi2016ndrangheta}, while highlighting a need to strengthen ties at the higher levels of the organizational pyramid for guaranteeing efficiency in coordinating illicit activities. Noticeably, as illustrated in the first panel of Figure~\ref{figure:6}, such an intensification of the underlying ties is inherently combined~with highly--structured security architectures that systematically obscure these increasingly--strong interactions while moving from the periphery of simple affiliates to the core of bosses. In fact, Figure~\ref{figure:6} provides clear evidence of security structures mostly in the within-- and across--block ties among the leading groups of bosses. 
To the best of our knowledge, these results provide the first quantitative illustration of how the conjectured efficiency--security tradeoff \citep{morselli2007efficiency} is realistically implemented in practice by organized crime to shape the structure of the observed network.

Notice that the combination of the two matrices in Figure~\ref{figure:6} yields an accurate reconstruction of the observed network $\bY$ in Figure~\ref{figure:7}. Moreover, besides incrementing knowledge on the structure and function of criminal networks, the ZIP--SBM also allows the identification of which currently--observed zero ties are more likely~to hide actual non--zero interactions and, as a consequence, should be object of further investigation by law enforcement. These guidelines can be obtained from the analysis of the posterior mean for the probabilities in \eqref{eq31}, which  is displayed in the second matrix of Figure~\ref{figure:7} and, consistent with the previous discussion, points toward the need of further investigations mainly on the across--block interactions among specific bosses from different {\em locali}. The reliability~of these results is supported by a low posterior uncertainty. For instance, the first, second and third quartiles of the posterior standard deviations for the entries of the matrices in Figure~\ref{figure:6} are  $[0.099,0.102, 0.107]$ and $[0.018, 0.031, 0.057]$, respectively.

\section{\Large 6. Conclusions and Future Research Directions}\label{sec_6}
This article develops innovative models and inference methods to answer a fundamental objective in criminology, namely the identification of redundancy, security and efficiency structures in organized crime from the analysis of weighted networks among the corresponding members. Despite the relevance of these architectures in guiding law enforcement \citep[see e.g.,][]{campana2022studying,diviak2022key}, the methodological challenges underlying the attempt to disentangle such structures have hindered advancements  along these lines. The proposed ZIP--SBM effectively covers this gap via a combination of stochastic block models, zero--inflated~Poisson distributions~and supervised Gnedin process priors which allow to incorporate fundamental concepts of redundancy and homophily, while crucially leveraging structure also in the observed zero ties to ultimately learn both security and efficiency structures. This yields a unique Bayesian modeling framework for criminal networks that can be implemented via effective data--augmentation collapsed Gibbs--samplers and provides superior performance along with innovative inference potentials relative to state--of--the--art alternatives, both in simulations and in applications. All these advantages are clear in the motivating {\em Infinito} network application, where the newly--proposed ZIP--SBM yields an unprecedented reconstruction of the security and efficiency structures behind a complex 'Ndrangheta organization from the analysis of weighted patterns of co--attendances to summits.

The aforementioned results are expected to stimulate further research along these directions. For example, from an applied perspective it would be of interest to consider more extensive implementations of the proposed ZIP--SBM to analyze currently--available criminal network data, such as those within the UCINET repository.~This would provide an important opportunity to understand  how security and efficiency structures vary~across different criminal organizations, including terrorists networks, or within the same organization but with respect to different forms of interaction, i.e., meetings, phone calls, family ties, and others, possibly monitored over time and/or by different investigation agencies. Such a multiplex and/or dynamic structure would also require extensions of the proposed ZIP--SBM  to allow both the block--specific parameters and, potentially, the partition structure, to vary across the different network views for the same criminal organization. The contributions by e.g., \citet{le2018,man2024, dur2014} and \citet{dur2017}  could provide valuable ideas and building--blocks to address this goal. From a methodological perspective it would be also of interest to consider alternative forms of zero--inflation in the general class of zero--inflated power series models \citep[e.g.,][]{ghosh2006bayesian} that may yield a more flexible characterization of weighted ties in other criminal networks, beyond those analyzed in this article.

\section{\bf Acknowledgments}
Daniele Durante is funded by the European Union (ERC, NEMESIS, project number: 101116718). Views and opinions expressed are however those of the author(s) only and do not necessarily reflect those of the European Union or the European Research Council Executive Agency. Neither the European Union nor the granting authority can be held responsible for them. 

Chaoyi Lu and Nial Friel are also affiliated to the Insight Centre for Data Analytics which is supported by~the Science Foundation Ireland under Grant Number 12/RC/2289$\_$P2.

The authors would like to thank Manuela Cattelan for the useful discussion and feedbacks on an initial version of the ZIP--SBM developed in this article.

\end{document}